\documentclass[12pt]{article}

\usepackage{amsmath}
\usepackage{amssymb}
\usepackage{latexsym}
\usepackage{graphicx}
\usepackage{epsfig}

\usepackage{tabularx} % extra features for tabular environment
\usepackage{amsmath}  % improve math presentation
\usepackage{graphicx} % takes care of graphic including machinery
\usepackage{cite} % takes care of citations
\usepackage[final]{hyperref} % adds hyper links inside the generated pdf file
\hypersetup{
	colorlinks=true,       % false: boxed links; true: colored links
	linkcolor=blue,        % color of internal links
	citecolor=blue,        % color of links to bibliography
	filecolor=magenta,     % color of file links
	urlcolor=blue         
	}

\def\beq{\begin{equation}}
\def\eeq{\end{equation}}
\def\bea{\begin{eqnarray}}
\def\eea{\end{eqnarray}}

\def\a{\alpha}

\def\m{\mu}
\def\n{\nu}

\def\f{\phi}
\def\L{\Lambda}

\def\nn{\nonumber}
\def\2{\;\;}
\def\4{\;\;\;\;}
\begin{document}

\begin{titlepage}

\vspace*{1cm}
\begin{center}
{\bf \Large Novel Black-Hole Solutions in \\[2mm]
Einstein-Scalar-Gauss-Bonnet Theories \\[4mm]
with a Cosmological Constant}

\bigskip \bigskip \medskip

{\bf A. Bakopoulos}$^{\,(a)}$\,\footnote{Email: abakop@cc.uoi.gr},
{\bf G. Antoniou}$^{\,(b)}$\,\footnote{Email: anton296@umn.edu}  and
{\bf P. Kanti}$^{\,(a)}$\,\footnote{Email: pkanti@cc.uoi.gr}

\bigskip
$^{(a)}${\it Division of Theoretical Physics, Department of Physics,\\
University of Ioannina, Ioannina GR-45110, Greece}

$^{(b)}${\it School of Physics and Astronomy, University of Minnesota,
Minneapolis, MN 55455, USA}

\bigskip \medskip
{\bf Abstract}
\end{center}
We consider the Einstein-scalar-Gauss-Bonnet theory in the presence of a cosmological
constant $\Lambda$, either positive or negative, and look for novel, regular black-hole
solutions with a non-trivial scalar hair. We first perform an analytic study in the near-horizon
asymptotic regime, and demonstrate that a regular black-hole horizon with a non-trivial
hair may be always formed, for either sign of $\Lambda$ and for arbitrary choices of the
coupling function between the scalar field and the Gauss-Bonnet term. At the far-away
regime, the sign of $\Lambda$ determines the form of the asymptotic gravitational
background leading either to a Schwarzschild-Anti-de Sitter-type background ($\Lambda<0$)
or a regular cosmological horizon ($\Lambda>0$), with a non-trivial scalar field in both
cases. We demonstrate that families of novel black-hole solutions with scalar hair emerge
for $\Lambda<0$, for every choice of the coupling function between the scalar field and
the Gauss-Bonnet term, whereas for $\Lambda>0$, no such solutions may be found. 
In the former case, we perform a comprehensive study of the physical properties of the
solutions found such as the temperature, entropy, horizon area and asymptotic behaviour
of the scalar field.

\end{titlepage}

\setcounter{page}{1}
%%%%%%%%%%%%%%%%%%%%%%%%%%%%%%%%%%%%%%%%%%%%%%%%%%%%%%%%%%%%%%%%%%%%%%

\section{Introduction}

As the ultimate theory of Quantum Gravity, that would robustly describe gravitational interactions
at high energies and facilitate their unification with the other forces, is still eluding us, the
interest in generalised gravitational theories remains unabated in the scientific literature.
These theories include extra fields or higher-curvature terms in their action \cite{Stelle,General},
and they provide the framework in the context of which several solutions of the traditional
General Relativity (GR) have been re-examined and, quite often, significantly enriched.

In this spirit, generalised gravitational theories containing scalar fields were among the
first to be studied. However, the quest for novel black-hole solutions -- beyond the three
well-known families of GR -- was abruptly stopped when the no-hair theorem was
formulated \cite{NH-scalar}, that forbade the existence of a static solution of this form
with a non-trivial scalar field associated with it. Nevertheless, counter-examples appeared
in the years that followed and included black holes with Yang-Mills \cite{YM}, Skyrme fields
\cite{Skyrmions} or with a conformal coupling to gravity \cite{Conformal}. A novel
formulation of the no-hair theorem was proposed in 1995 \cite{Bekenstein} but
this was, too, evaded within a year with the discovery of the dilatonic black holes
found in the context of the Einstein-Dilaton-Gauss-Bonnet theory \cite{DBH}
(for some earlier studies that paved the way, see \cite{Gibbons, Callan, Campbell,
Mignemi, Kanti1995}). The coloured black holes were found next in the context of the
same theory completed by the presence of a Yang-Mills field \cite{Torii, KT}, and
higher-dimensional \cite{Guo} or rotating versions \cite{Kleihaus, Pani, Herdeiro, Ayzenberg}
were also constructed (for a number of interesting reviews on the topic, see
\cite{Win-review, Charmousis-rev, Herdeiro-review, Blazquez}).

This second wave of black-hole solutions were derived in the context of theories inspired
by superstring theory \cite{Metsaev}. During the last decade, though, the construction
of generalised gravitational theories was significantly enlarged via the revival of the 
Horndeski \cite{Horndeski} and Galileon \cite{Galileon} theories. Accordingly, novel
formulations of the no-hair theorems were proposed that covered the case of standard
scalar-tensor theories \cite{SF} and Galileon fields \cite{HN}. However, these recent forms
were also evaded \cite{SZ} and concrete black-hole solutions were constructed
\cite{Babichev, Benkel, Yunes2011}. More recently, three independent groups \cite{ABK, Doneva, Silva}
almost simultaneously demonstrated that a generalised gravitational theory that contains
a scalar field and the quadratic Gauss-Bonnet (GB) term admits novel black-hole solutions
with a non-trivial scalar hair. In a general theoretical argument, that we presented in
\cite{ABK}, it was shown that the presence of the GB term was of paramount importance
for the evasion of the novel no-hair theorem \cite{Bekenstein}. In addition, the exact form
of the coupling function $f(\phi)$ between the scalar field and the GB term played no
significant role for the emergence of the solutions: as long as the first derivative of the
scalar field $\phi_h'$ at the horizon obeyed a specific constraint, an asymptotic solution
describing a regular black-hole horizon with a non-trivial scalar field could always be
constructed. Employing, then, several different forms of the coupling function $f(\phi)$,
a large number of asymptotically-flat black-hole solutions with scalar hair were determined
\cite{ABK}. Additional studies presenting novel black holes or compact objects in 
generalised gravitational theories have appeared \cite{Charmousis, Correa, Doneva-NS,
Motohashi, Radu, Doneva-Papa, Butler, Danila, Stetsko} as well as further studies of the properties
of these novel solutions \cite{Ayzen, Dolan, Kunz, Bhatta, Tatter, Mukherjee, Chakra, Berti,
Brihaye, Prabhu, Myung, Don-Kunz, Benkel2018, Iorio, Ovalle, Barack, Gao, Lee, Witek2018, Moto}. 

In the present work, we will extend our previous analyses \cite{ABK}, that aimed at deriving
asymptotically-flat black-hole solutions, by introducing in our theory a cosmological
constant $\Lambda$, either positive or negative. In the context of this theory, we will
investigate whether the previous, successful synergy between the Ricci scalar, the scalar
field and the Gauss-Bonnet term survives in the presence of $\Lambda$. The question of
the existence of black-hole solutions in the context of a scalar-tensor theory, 
with scalar fields minimally-coupled or conformally-coupled to gravity, and a 
cosmological constant has been debated in the literature for decades
\cite{Narita, Winstanley-Nohair, Bhatta2007, Bardoux, Kazanas}. 
In the case of a {\it positive} cosmological constant, the existing studies predominantly
excluded the presence of a regular, black-hole solution with an asymptotic de Sitter
behaviour - a counterexample of a black hole in the context of a theory with a
conformally-coupled scalar field \cite{Martinez-deSitter} was shown later to be unstable 
\cite{Harper}. On the other hand, in the case of a {\it negative} cosmological constant,
a substantial number of solutions with an asymptotically AdS behaviour have been found
in the literature (for a non-exhaustive list, see \cite{Martinez, Radu-Win, Anabalon, Hosler,
Kolyvaris, Ohta, Saenz, Caldarelli, Gonzalez, Gaete, Giribet, BenAchour}. 

Here, we perform a comprehensive study of the existence of black-hole solutions
with a non-trivial scalar hair and an asymptotically (Anti)-de Sitter behaviour in 
the context of a general class of theories containing the higher-derivative, quadratic
GB term. To our knowledge, the only similar study is the one performed in the special
case of the shift-symmetric Galileon theory \cite{Hartmann}, i.e. with a linear coupling
function between the scalar field and the GB term. In this work, we consider the most general
class of this theory by considering an arbitrary form of the coupling function $f(\phi)$,
and look for regular black-hole solutions with non-trivial scalar hair. Since the uniform
distribution of energy associated with the cosmological constant permeates the whole
spacetime, we expect $\Lambda$ to have an effect on both the near-horizon and
far-field asymptotic solutions. We will thus repeat our analytical calculations both in
the small and large-$r$ regimes to examine how the presence of $\Lambda$ affects
the asymptotic solutions both near and far away from the black-hole horizon. As we
will see, our set of field equations admits regular solutions near the black-hole horizon
with a non-trivial scalar hair for both signs of the cosmological constant. At the 
far-away regime, the analysis needs to be specialised since a positive or negative
sign of $\Lambda$ leads to either a cosmological horizon or an asymptotic
Schwarzschild-Anti-de Sitter-type
gravitational background, respectively. Our results show that the emergence of a black-hole
solution with a non-trivial scalar hair strongly depends on the type of asymptotic background
that is formed at large distances, and thus on the sign of $\Lambda$: whereas, for
$\Lambda<0$, solutions emerge with the same easiness as their asymptotically-flat
analogues, for $\Lambda>0$, no such solutions were found. 

In the former case, i.e. for $\Lambda<0$, we present a large number of novel black-hole
solutions with a regular black-hole horizon, a non-trivial scalar field and a
Schwarzschild-Anti-de Sitter-type asymptotic behaviour at large distances, These solutions
correspond to a variety of forms of the coupling function $f(\phi)$: exponential, polynomial
(even or odd), inverse polynomial (even or odd) and logarithmic. Then, we proceed to study
their physical properties such as the temperature, entropy, and horizon area. We also
investigate features of the asymptotic profile of the scalar field, namely its effective potential
and rate of change at large distances since this greatly differs from the asymptotically-flat case.

The outline of the present work is as follows: in Section 2, we present our theoretical framework
and perform our analytic study of the near and far-way radial regimes as well as of their
thermodynamical properties. In Section 3, we present our numerical results for the two cases
of $\Lambda<0$ and $\Lambda>0$. We finish with our conclusions in Section 4.

%%%%%%%%%%%%%%%%%%%%%%%%%%%%%%%%%%%%%%%%%%%%%%%%%%%%%%%%%%%%%%%%%%5

\section{The Theoretical Framework}

We consider a general class of higher-curvature gravitational theories 
described by the following action functional:
%%%%%%%%%%%%%%%%%%%%
\begin{equation}
S=\frac{1}{16\pi}\int{d^4x \sqrt{-g}\left[R-\frac{1}{2}\,\partial_{\mu}\phi\partial^{\mu}\phi+f(\phi)R^2_{GB}- 2\L\right]}.
\label{action}
\end{equation}
%%%%%%%%%%%%%% 
In this, the quadratic Gauss-Bonnet (GB) term $R^2_{GB}$, defined as 
%%%%%%%%%%%%
\begin{equation}\label{GB-def}
R^2_{GB}=R_{\mu\nu\rho\sigma}R^{\mu\nu\rho\sigma}-4R_{\mu\nu}R^{\mu\nu}+R^2\,,
\end{equation}
%%%%%%%%%%%%
supplements the Einstein-Hilbert term, given by the Ricci scalar curvature $R$, and
the kinetic term for a scalar field $\phi$. A coupling term of the scalar field
to the GB term, through a general coupling function $f(\phi)$, is necessary 
in order for the GB term -- a total derivative in four dimensions -- to
contribute to the field equations. A cosmological constant $\Lambda$, that
may take either a positive or a negative value, is also present in the theory.

By varying the action (\ref{action}) with respect to the metric tensor $g_{\mu\nu}$
and the scalar field $\phi$, we derive the gravitational field equations and the
equation for the scalar field, respectively. These are found to have the form:
%%%%%%%%%%%%%%%%
\begin{equation}
G_{\mu\nu}=T_{\mu\nu}\,, \label{field-eqs}
\end{equation}
\begin{equation}
\nabla^2 \phi+\dot{f}(\phi)R^2_{GB}=0\,, \label{phi-eq_0}
\end{equation}
%%%%%%%%%%%%%%
where $G_{\mu\nu}$ is the Einstein tensor and $T_{\mu\nu}$ is the energy-momentum tensor,
with the latter having the form
%%%%%%%%%%%%%%%%%%%
\begin{equation}\label{Tmn}
T_{\mu\nu}=-\frac{1}{4}g_{\mu\nu}\partial_{\rho}\phi\partial^{\rho}\phi+\frac{1}{2}\partial_{\mu}\phi\partial_{\nu}\phi-\frac{1}{2}\left(g_{\rho\mu}g_{\lambda\nu}+g_{\lambda\mu}g_{\rho\nu}\right)
\eta^{\kappa\lambda\alpha\beta}\tilde{R}^{\rho\gamma}_{\quad\alpha\beta}
\nabla_{\gamma}\partial_{\kappa}f(\phi)- \L\,g_{\m\n}\,.
\end{equation}
%%%%%%%%%%%%%
In the above, the dot over the coupling function denotes its derivative with respect
to the scalar field (i.e. $\dot f =df/d\phi$). We have also employed units in which $G=c=1$,
and used the definition
%%%%%%%%%%%%%%%%
\begin{equation}\label{tildeR}
\tilde{R}^{\rho\gamma}_{\quad\alpha\beta}=\eta^{\rho\gamma\sigma\tau}
R_{\sigma\tau\alpha\beta}=\frac{\epsilon^{\rho\gamma\sigma\tau}}{\sqrt{-g}}\,
R_{\sigma\tau\alpha\beta}\,.
\end{equation}
%%%%%%%%%%%%%%%%%%%%%%%%
Compared to the theory studied in \cite{ABK}, where $\Lambda$ was zero, the changes
in Eqs. (\ref{field-eqs})-(\ref{phi-eq_0}) look minimal: the scalar-field equation
remains unaffected while the energy-momentum tensor $T^\mu_{\;\, \nu}$ receives a
constant contribution $-\Lambda \delta^{\mu}_{\;\, \nu}$.
However, as we will see, the presence of the cosmological constant affects both of the
asymptotic solutions, the properties of the derived black holes and even their existence.

In the context of this work, we will investigate the emergence of regular, static,
spherically-symmetric but non-asymptotically flat black-hole solutions with a non-trivial
scalar field. The line-element of space-time will accordingly take the form
%%%%%%%%%%%%%%%%
\begin{equation}\label{metric}
{ds}^2=-e^{A(r)}{dt}^2+e^{B(r)}{dr}^2+r^2({d\theta}^2+\sin^2\theta\,{d\varphi}^2)\,.
\end{equation}
%%%%%%%%%%%%%%%
The scalar field will also be assumed to be static and spherically-symmetric, $\phi=\phi(r)$.
The coupling function $f(\phi)$ will retain a general form during the first part
of our analysis, and will be chosen to have a particular form only at the stage of
the numerical derivation of specific solutions.

The non-vanishing components of the Einstein tensor $G^\mu_{\;\, \nu}$ may be easily
found by employing the line-element (\ref{metric}), and they read
%%%%%%%%%%%%%
\begin{align}
 G^t_{\;\, t} &= \frac{e^{-B}}{r^2}(1-e^B-rB'),\label{Gtt}\\
 G^r_{\;\, r} &= \frac{e^{-B}}{r^2}(1-e^B+rA'),\label{Grr}\\
 G^\theta_{\;\,\theta} &=G^\phi_{\;\,\phi}=
 \frac{e^{-B}}{4r}\left[r{A'}^2-2B'+A'(2-rB')+2rA''\right].\label{Gthth}
\end{align}
%%%%%%%%%%%%%%%
Throughout our analysis, the prime denotes differentiation with respect to the radial
coordinate $r$. Using Eq. (\ref{Tmn}), the components of the energy-momentum tensor
$T^\mu_{\;\, \nu}$ take in turn the form 
%%%%%%%%%%%%%%%%%%%%
\begin{align}
T^t_{\;\,t}=&-\frac{e^{-2B}}{4r^2}\left[\phi'^2\left(r^2e^B+16\ddot{f}(e^B-1)\right)-8\dot{f}\left(B'\phi'(e^B-3)-2\phi''(e^B-1)\right)\right]-\L, \label{Ttt}\\[0mm]
T^r_{\;\,r}=&\frac{e^{-B}\phi'}{4}\left[\phi'-\frac{8e^{-B}\left(e^B-3\right)\dot{f}A'}{r^2}\right] -\L, \label{Trr}\\[2mm]
T^{\theta}_{\;\,\theta}=&T^{\varphi}_{\;\,\varphi}=-\frac{e^{-2B}}{4 r}\left[\phi'^2\left(re^B-8\ddot{f}A'\right)-4\dot{f}\left(A'^2\phi'+2\phi'A''+A'(2\phi''-3B'\phi')\right)\right]
-\L. \label{Tthth}
\end{align}
%%%%%%%%%%%%%%%%%%
Matching the corresponding components of $G^\mu_{\;\, \nu}$ and $T^\mu_{\;\, \nu}$,
the explicit form of Einstein's field equations may be easily derived. These are supplemented
by the scalar-field equation (\ref{phi-eq_0}) whose explicit form reads
%%%%%%%%%%%%%%%%%%%
\begin{equation}
2r\phi''+(4+rA'-rB')\,\phi'+\frac{4\dot{f}e^{-B}}{r}\left[(e^B-3)A'B'-(e^B-1)(2A''+A'^2)\right]=0\,. \label{phi-eq}
\end{equation}
%%%%%%%%%%%%%%%%%%%%%

Although the system of equations involve three unknown functions, namely $A(r)$, $B(r)$ and
$\phi(r)$, only two of them are independent. The metric function $B(r)$ may be easily shown
to be a dependent variable: the $(rr)$-component of field equations takes in fact the form of
a second-order polynomial with respect to $e^B$, i.e. $\alpha e^{2B}+\beta e^{B}+\gamma=0$,
which easily leads to the following solution 
%%%%%%%%%%%%%%%
\begin{equation}\label{Bfunction}
e^B=\frac{-\beta\pm\sqrt{\beta^2-4\a\gamma}}{2\a},
\end{equation}
where
\beq
\a= 1-\L r^2, \qquad
\beta=\frac{r^2{\phi'}^2}{4}-(2\dot{f}\phi'+r) A'-1,\label{abc}\qquad
\gamma=6\dot{f}\phi'A'.%\label{gamma}
\eeq
%%%%%%%%%%%%%%%%%
Employing the above expression for $e^B$, the quantity $B'$ may be also found to have the form
%%%%%%%%%%%%%%%%
\begin{equation}\label{B'}
B'=-\frac{\gamma'+\beta' e^{B}+\a' e^{2B}}{2\a e^{2B}+\beta e^{B}}.
\end{equation}
%%%%%%%%%%%%%
Therefore, by using Eqs. (\ref{Bfunction}) and (\ref{B'}), the metric function $B(r)$ may be
completely eliminated from the field equations. The remaining three equations then form a
system of only two independent, ordinary differential equations of second order for
the functions $A(r)$ and $\phi(r)$:
%%%%%%%%%%%%%%%%%%%%%
\begin{align}
A''=&\frac{P}{S}\,,  \label{A-sys} \\
\phi''=&\frac{Q}{S}\,. \label{phi}
\end{align} 
The expressions for the quantities $P$, $Q$ and $S$, in terms of $(r, \phi', A', \dot f, \ddot f)$,
are given for the interested reader in Appendix A as they are quite complicated. 

\subsection{Asymptotic Solution at Black-Hole Horizon}

As we are interested in deriving novel black-hole solutions, we will first investigate
whether an asymptotic solution describing a regular black-hole horizon is admitted
by the field equations. As a matter of fact, instead of assuming the usual power-series
expression in terms of $(r-r_h)$, where $r_h$ is the horizon radius, we will construct
the solution as was done in \cite{DBH, ABK}. To this end, we demand that,
near the horizon, the metric function $e^{A(r)}$ should vanish (and $e^{B(r)}$ should
diverge) whereas the scalar field must remain finite. The first demand is reflected in the
assumption that $A'(r)$ should diverge as $r \rightarrow r_h$ -- this will be
justified {\it a posteriori} -- while $\phi'(r)$ and $\phi''(r)$ must be finite in the
same limit. 

Assuming the aforementioned behaviour near the black-hole horizon, Eq. (\ref{Bfunction})
may be expanded in terms of $A'(r)$ as follows\footnote{Note, that only the (+)-sign
in the expression for $e^B$ in Eq. (\ref{Bfunction}) leads to the desired black-hole
behaviour.}
%%%%%%%%%%%%%%%%
\bea\label{expb}
e^B=\frac{(2\dot f \phi '+ r)}{1-\Lambda  r^2}\,A'  - 
\frac{2 \dot f \phi ' \left(r^2 \phi '^2-12 \Lambda  r^2+8\right)+r \left(r^2 \phi '^2-4\right)}
{4 (1-\Lambda r^2)\,(2 \dot f \phi '+r)} + \mathcal{O}\left(\frac{1}{A'}\right).
\eea
%%%%%%%%%%%%%%
Then, substituting the above into the system (\ref{A-sys})-(\ref{phi}), we obtain 
\begin{align}
A''=&\frac{W_1}{W_3}\,A'^2+\mathcal{O}\left(A'\right),\label{expa}\\[3mm]
\f''=&\frac{W_2}{W_3}\,(2 \dot f \phi '+r) A' + \mathcal{O}(1),\label{expf}
\end{align}
where
\bea\label{w1}
W_1&=& -(r^4+4 r^3 \dot f \phi' + 4 r^2 \dot f^2 \phi '^2-24 \dot f^2) + 
24 \Lambda ^2 r^4 \dot f^2 \nn\\[2mm]
   &&+   \Lambda \left[4 r^5 \dot f \phi'+4 r^2 \dot f^2 \left(r^2 \phi'^2-16\right)
   -64 r \dot f^3 \phi '-64 \dot f^4 \phi'^2+r^6\right], 
\eea
\bea\label{w2}
W_2&=& -r^3 \phi ' \left(1-\Lambda  r^2\right) -32 \Lambda  \dot f^3 \phi '^2+
       16 \Lambda  r \dot f ^2 \phi ' \left(\Lambda  r^2-3\right)\nn\\[2mm]
  && -2 \dot f \left[6+ r^2\phi '^2 + 2 \Lambda ^2 r^4-\Lambda  r^2 \left(r^2 \phi'^2+4\right)\right],
\eea
%%%%%%%%%%%%%%
and
%%%%%%%%%%%%%%
\beq
W_3=\left(1-\Lambda  r^2 \right)
   \left[r^4 + 2 r^3 \dot f \phi '-16 \dot f^2 \left(3-2 \Lambda  r^2\right)-
    32 \Lambda  r \dot f^3 \phi '\right].\label{w3}
\eeq
%
%%%%%%%%%%%%%%%%
From Eq. (\ref{expb}), we conclude that the combination $(2\dot f \phi '+ r)$ near the horizon
must be non-zero and positive for the metric function $e^B$ to have the correct behaviour,
that is to diverge as $r \rightarrow r_h$ while being positive-definite. Then, Eq. (\ref{expf})
dictates that, if we want $\phi''$ to be finite, we must necessarily have 
%%%%%%%%%%%%%%%
\begin{equation}
W_2|_{r=r_h}=0\,.
\end{equation}
%%%%%%%%%%%%%%%%
The above constraint may be written as a second-order polynomial with respect to $\phi'$,
which can then be solved to yield
%%%%%%%%%%%%%%%%
\begin{align}\label{solf}
\f'_h=-\frac{r_h^3 (1- \L r_h^2)+16 \Lambda  r_h \dot f^2_h (3-\Lambda  r_h^2)
      \pm (1-\Lambda  r_h^2)\sqrt{C}}{4 \dot f \,\Bigl[r_h^2 -\Lambda (r_h^4-16
   \dot f^2_h)\Bigr]},
\end{align}
%%%%%%%%%%%%%%%%
where all quantities have been evaluated at $r=r_h$ The quantity $C$ under the square root
stands for the following combination
\begin{equation}
C=256 \Lambda  \dot f^4_h \left(\Lambda  r_h^2-6\right) +
   32 r_h^2 \dot f^2_h \left(2\Lambda  r_h^2-3\right)+r_h^6 \geq 0\,,
\label{C-def}
\end{equation}
%%%%%%%%%%%%%%
and must always be non-negative for $\phi'_h$ to be real. This combination may be written
as a second-order polynomial for $\dot f^2_h$ with roots
%%%%%%%%%%%%%%%%%%
\beq 
\dot f^2_{\pm}=\frac{r_h^2 \left[3-2 \Lambda r_h^2 \pm \sqrt{3}\sqrt{3-2 \Lambda r_h^2 +
\Lambda^2 r_h^4}\right]}{16 \Lambda\,(-6 + \Lambda r_h^2)}\,.
\label{con-C}
\eeq
%%%%%%%%%%%%%%%%
Then, the constraint on $C$ becomes
%%%%%%%%%%%%%
\beq
C=(\dot f^2_h -\dot f^2_{-})\,(\dot f^2_h -\dot f^2_{+}) \geq 0\,.
\label{C-newdef}
\eeq
%%%%%%%%%%%
Therefore, the allowed regime for the existence of regular, black-hole solutions with 
scalar hair is given by $\dot f^2_h \leq \dot f^2_{-}$ or $\dot f^2_h \geq \dot f^2_{+}$,
since $\dot f^2_{+}>\dot f^2_{-}$. To obtain some physical insight on these inequalities,
we take the limit of small cosmological constant; then, the allowed ranges are
%%%%%%%%%%%%%%
\beq
\dot f^2_h \leq  \frac{r_h^4}{96} \left(1 + \frac{\Lambda r_h^2}{6} + ... \right),
\qquad {\rm or,} \qquad
\dot f^2_h \geq  \frac{r_h^4}{48} \left(1 - \frac{3}{\Lambda r_h^2} + ... \right),
\label{f-ineq}
\eeq
%%%%%%%%%%%%%%%%%%%
respectively. In the absence of $\Lambda$, Eq. (\ref{C-def}) results into the
simple constraint $\dot f^2_h \leq r_h^4/96$, and defines a sole branch of solutions with a
minimum allowed value for the horizon radius (and mass) of the black hole \cite{ABK}.
In the presence of a cosmological constant, this constraint is now replaced by 
$\dot f^2_h \leq \dot f^2_{-}$, or by the first inequality presented in Eq. (\ref{f-ineq})
in the small-$\Lambda$ limit. This inequality leads again to a branch of solutions that --
for chosen $f(\phi)$, $\phi_h$ and $\Lambda$ -- terminates at a 
black-hole solution with a minimum horizon radius $r_{h}^{min}$. We observe that, at least for
small values of $\Lambda$, the presence of a positive cosmological constant relaxes the constraint
allowing for smaller black-hole solutions, while a negative cosmological constant pushes the
minimum horizon radius towards larger values. The second inequality in Eq. (\ref{f-ineq})
describes a new branch of black-hole solutions that does not exist when $\Lambda=0$;
this was also noted in \cite{Hartmann} in the case of the linear coupling function. This branch
of solutions describes a class of very small GB black holes, and terminates instead at a
black hole with a maximum horizon radius $r_{h}^{max}$. 

Returning now to Eq.  (\ref{A-sys}) and employing the constraint (\ref{solf}), the former takes the form
%%%%%%%%%%%%%
\beq
A''=-A'^2+\mathcal{O}\left(A'\right).\label{expa2}
\eeq
%%%%%%%%%%%%%%%
Integrating the above, we find that $A'(r) \sim 1/(r-r_h)$, a result that justifies the diverging
behaviour of this quantity near the horizon that we assumed earlier. A second integration
yields  $A(r) \sim \ln (r-r_h)$, which then uniquely determines the expression of the metric
function $e^A$ in the near-horizon regime. Employing Eq. (\ref{expb}), the metric
function $B$ is also determined in the same regime. Therefore, the asymptotic solution
of Eqs. (\ref{Bfunction}), (\ref{A-sys}) and (\ref{phi}), that describes a regular, black-hole
horizon in the limit $r \rightarrow r_h$, is given by the following expressions
%%%%%%%%%%%%%
\bea
&&e^{A}=a_1 (r-r_h) + ... \,, \label{A-rh}\\[1mm]
&&e^{-B}=b_1 (r-r_h) + ... \,, \label{B-rh}\\[1mm]
&&\phi =\phi_h + \phi_h'(r-r_h)+ \phi_h'' (r-r_h)^2+ ... \,, \label{phi-rh}
\eea
%%%%%%%%%%%%%
where $a_1, b_1$ and $\phi_h$ are integration constants.
We observe that the above asymptotic solution constructed for the case of a non-zero
cosmological constant has exactly the same functional form as the one constructed
in \cite{ABK} for the case of vanishing $\Lambda$. The presence of the cosmological
constant modifies though the exact expressions of the basic constraint (\ref{solf})
for $\phi'_h$ and of the quantity $C$ given in (\ref{C-def}), the validity of which 
ensures the existence of a regular black-hole horizon. As in \cite{ABK}, the
exact form of the coupling function $f(\phi)$ does not affect the existence of the
asymptotic solution, therefore regular black-hole solutions may emerge for a wide
class of theories of the form (\ref{action}). 

The regularity of the asymptotic black-hole solution is also reflected in the non-diverging
behaviour of the components of the energy-momentum tensor and of the scale-invariant
Gauss-Bonnet term. The components of the former quantity in this regime assume the form
%%%%%%%%%%%%%%%%%%%
\begin{align}
T^t_{\;\,t}&=\frac{2e^{-B}}{r^2}\,B'\f' \dot f - \L +\mathcal{O}(r-r_h),\label{Ttt_rh}\\[3mm]
T^r_{\;\,r}&=-\frac{2e^{-B}}{r^2}\,A'\f' \dot f - \L+\mathcal{O}(r-r_h),\label{Trr_rh}\\[3mm]
T^\theta_{\;\,\theta}&=\frac{e^{-2B}}{r}\,(2A'' +A'^2-3A'B')\,\f' \dot f - \L+\mathcal{O}(r-r_h).
\label{Tthth_rh}
\end{align}
%%%%%%%%%%%%%
Employing the asymptotic expansions (\ref{A-rh})-(\ref{phi-rh}), one may see that all
components remain indeed finite in the vicinity of the black-hole horizon.
For future use, we note that the cosmological constant adds a positive contribution to 
all components of the energy-momentum tensor $T^\mu_{\;\,\nu}$ for $\Lambda<0$, while it
subtracts a positive contribution for $\Lambda>0$. Also, all scalar curvature quantities,
the explicit form of which may be found in Appendix B, independently exhibit a regular
behaviour near the black-hole horizon -- when these are combined, the GB term, in
the same regime, takes the form
%%%%%%%%%% 
\beq
R^2_{GB} = +\frac{12  e^{-2 B}}{r^2}A'^2 +\mathcal{O}(r-r_h)\,, \label{GB-rh}
\eeq
%%%%%%%%%%%%%
%\beq
%R^2_{GB}\rightarrow\frac{12 b_1^2}{r^2} \label{GB_rh}
%\eeq
%%%%%%%%%%%%%%
exhibiting, too, a regular behaviour as expected.

%%%%%%%%%%%%%%%%%%%%%%%%%%%%%%%%%%%%%%%%%%%%%%%%%%%%%%%%%%%%%%%%%

\subsection{Asymptotic Solutions at Large Distances}

The form of the asymptotic solution of the field equations at large distances from the
black-hole horizon depends strongly on the sign of the cosmological constant. Therefore,
in what follows, we study separately the cases of positive and negative $\Lambda$. 

%%%%%%%%%%%%%%%%%%%%%%%%%%%%%%%%%%%%%

\subsubsection{Positive Cosmological Constant}

In the presence of a positive cosmological constant, a second horizon, the cosmological one, 
is expected to emerge at a radial distance $r=r_c>r_h$. We demand that this horizon is
also regular, that is that the scalar field $\phi$ and its derivatives remain finite in its
vicinity. We may in fact follow a method identical to the one followed in section 2.1 near
the black-hole horizon: we again demand that, at the cosmological horizon, $g_{tt} \rightarrow 0$
while $g_{rr} \rightarrow \infty$; then, using that $A'$ diverges there, the regularity of $\phi''$
from Eq. (\ref{phi}) eventually leads to the constraint 
%%%%%%%%%%%%%%%%
\begin{align}\label{solfc}
\f'_c=-\frac{r_c^3 (1- \L r_c^2)+16 \Lambda  r_c \dot f^2_c (3-\Lambda  r_c^2)
      \pm (1-\Lambda  r_c^2)\sqrt{\tilde C}}{4 \dot f \,\Bigl[r_c^2 -\Lambda (r_c^4-16
   \dot f^2_c)\Bigr]},
\end{align}
with $\tilde C$ now being given by the non-negative expression
\begin{equation}\label{tildeC-def}
\tilde C=256 \Lambda  \dot f^4_c \left(\Lambda  r_c^2-6\right) +
   32 r_c^2 \dot f^2_c \left(2\Lambda  r_c^2-3\right)+r_c^6 \geq0 \,.
\end{equation} 
%%%%%%%%%%%%%%
Employing Eq. (\ref{solfc}) in Eq. (\ref{A-sys}), the solution for the metric function $A$ may
be again constructed. Overall, the asymptotic solution of the field equations near a regular,
cosmological horizon will have the form
%%%%%%%%%%%%%%%%%%%%
\begin{align}
e^A&=a_2\,(r_c-r)+... , \label{A-rc}\\[3mm]
e^{-B}&=b_2\,(r_c-r)+... , \label{B-rc}\\[3mm]
\f&=\f_c+\f_c'(r_c-r)+\f_c''(r_c-r)^2+... , \label{phi-rc}
\end{align}
%%%%%%%%%%%%%
where care has been taken for the fact that $r \leq r_c$. One may see again that
the above asymptotic expressions lead to finite values for the components of the energy-momentum
tensor and scalar invariant quantities. Once again, the explicit form of the coupling function
$f(\phi)$ is of minor importance for the existence of a regular, cosmological horizon.

\subsubsection{Negative Cosmological Constant}

For a negative cosmological constant, and at large distances from the black-hole horizon,
we expect the spacetime to assume a form close to that of the Schwarzschild-Anti-de Sitter
solution. Thus, we assume the following approximate forms for the metric functions
%%%%%%%%%%%%%%%%%%
\bea
e^{A(r)}&=& \left(k-\frac{2M}{r}-\frac{\L_{eff}}{3}\,r^2+\frac{q_2}{r^2}\right)
\left(1+\frac{q_1}{r^2}\right)^2,\label{alfar1}\\[3mm]
e^{-B(r)}&=& k-\frac{2M}{r}-\frac{\L_{eff}}{3}\,r^2 +\frac{q_2}{r^2},\label{bfar1}
\eea
%%%%%%%%%%%%%%%%%%
where $k$, $M$, $\Lambda_{eff}$ and $q_{1,2}$ are, at the moment, arbitrary constants. 
%%%%%%%%%%%%%
Substituting the above expressions into the scalar field equation (\ref{phi-eq}), we obtain at first
order the constraint
%%%%%%%%%%%%%%%
\beq
\phi''(r)+\frac{4}{r}\,\phi'(r) - \frac{8 \Lambda_{eff} \dot f}{r^2}=0\,.
\label{phi-Anti-far}
\eeq
%%%%%%%%%%%%
The gravitational equations, under the same assumptions, lead to two additional
constraints, namely
%%%%%%%%%%%%%%%%
\beq
\Lambda - \Lambda_{eff} +\frac{\Lambda_{eff}\,r^2  \phi'}{12}\left(\phi'-
\frac{16 \Lambda_{eff} \dot f}{r}\right)=0, \label{con-far-1}
\eeq
%%%%%%%%%%%
\beq
\Lambda - \Lambda_{eff} -\frac{4}{9}\,\dot f \Lambda_{eff}^2 r^2 \left(\phi'' +
\frac{3 \phi'}{r}\right) -\frac{\Lambda_{eff}\,r^2}{12}\,\phi'^2 \left(1+
\frac{16 \Lambda_{eff} \ddot f}{3}\right)=0\,.
\label{con-far-2}
\eeq
%%%%%%%%%%%%%

Contrary to what happens close to the horizons (either black-hole or cosmological ones),
the form of the coupling function $f(\phi)$ now affects the asymptotic form of the scalar
field at large distances. The easiest case is that of a linear coupling function,
$f(\phi)=\alpha \phi$ - that case was first studied in \cite{Hartmann}, however, we
review it again in the context of our analysis as it will prove to play a more general
role.  The scalar field, at large distances, may be shown to have the approximate form
%%%%%%%%%%%%%%%%
\beq
\phi(r) = \phi_\infty + d_1 \ln r +\frac{d_2}{r^2} +  \frac{d_3}{r^3}+ ...\,, \label{phi-far-Anti}
\eeq
%%%%%%%%%%%%%%
where again $(\phi_\infty, d_1, d_2, d_3)$ are arbitrary constant coefficients. The coefficients
$d_1$ and $\Lambda_{eff}$ may be determined through the first-order constraints 
(\ref{phi-Anti-far}) and (\ref{con-far-1}), respectively, and are given by
%%%%%%%%%%%%%%
\beq
d_1=\frac{8}{3}\,\alpha \Lambda_{eff}\,, \qquad 
\Lambda_{eff} \left( 3 +\frac{80 \alpha^2 \Lambda_{eff}^2}{9}\right)=3 \Lambda\,.
\label{d1-Leff}
\eeq
%%%%%%%%%%%%%%%%%
The third first-order constraint, Eq. (\ref{con-far-2}), is then trivially satisfied. 
In order to determine the values of the remaining coefficients, one needs to derive
higher-order constraints. For example, the coefficients $k$, $q_1$ and $d_2$ are found
at third-order approximation to have the forms
%%%%%%%%%%%%%%%%
\beq
k=\frac{81 + 864\,\alpha^2 \Lambda_{eff}^2 + 1024\,\alpha^4 \Lambda_{eff}^4}
{81 + 1008\,\alpha^2 \Lambda_{eff}^2 + 2560\,\alpha^4 \Lambda_{eff}^4}\,,
\qquad 
q_1=\frac{24\,\alpha^2 \Lambda_{eff}\,(9+ 64\,\alpha^2 \Lambda_{eff}^2)}
{(9 + 32\,\alpha^2 \Lambda_{eff}^2)\,(9 + 80\,\alpha^2 \Lambda_{eff}^2)}\,,
\nonumber
\eeq
%%%%%%%%%%%%
\beq
d_2=-\frac{12 \alpha\,(27 + 288\,\alpha^2 \Lambda_{eff}^2 + 512\,\alpha^4 \Lambda_{eff}^4)}
{81 + 1008\,\alpha^2 \Lambda_{eff}^2 + 2560\,\alpha^2 \Lambda_{eff}^2}\,,
\label{k-q1}
\eeq
%%%%%%%%%%%%%%%%%%%
while for $q_2$ or $d_3$ one needs to go even higher. In contrast, the coefficient $M$ remains
arbitrary and may be interpreted as the gravitational mass of the solution.

In the perturbative limit (i.e. for small values of the coupling constant $\alpha$
of the GB term), one may show that the above asymptotic solution is valid for
all forms of the coupling function $f(\phi)$. Indeed, if we write
%%%%%%%%%%%%%%%%
\beq
\phi(r)= \phi_0 +\sum_{n=1}^\infty \alpha^n\,\phi_n(r)\,,
\label{phi-pert}
\eeq
%%%%%%%%%%%%%%%%%
and define $f(\phi)=\alpha \tilde f(\phi)$, then, at first order, 
$\dot f \simeq \alpha \,\dot{\tilde f}(\phi_0)$.
Therefore, independently of the form of $f(\phi)$, at first order in the perturbative limit, 
$\dot f$ is a constant, as in the case of a linear coupling function. Then, a solution
of the form of Eqs. (\ref{alfar1})-(\ref{bfar1}) and (\ref{phi-far-Anti}) is easily 
derived~\footnote{In the perturbative limit, at first order, one finds $d_1=8 \Lambda \dot f(\phi_0)/3$, $\Lambda_{eff}=\Lambda$, $k=1$, $q_1=0$, and $d_2=-4 \dot f(\phi_0)$.
For more details on the perturbative analysis of the black-hole solutions that arise in the
context of the general class of theories (\ref{action}) and are either asymptotically-flat or
(Anti)-de Sitter, see \cite{BAK2}.} with $\alpha$ in Eqs. (\ref{d1-Leff}) and (\ref{k-q1}) being
now replaced by $\dot f(\phi_0)$.

For arbitrary values of the coupling constant $\alpha$, though, or for a non-linear
coupling function $f(\phi)$, the approximate solution described by Eqs. (\ref{alfar1}),
(\ref{bfar1}) and (\ref{phi-far-Anti}) will not, in principle, be valid any more. Unfortunately,
no analytic form of the solution at large distances may be derived in these cases. However,
as we will see in section 3, numerical solutions do emerge with a non-trivial scalar field
and an asymptotic Anti-de Sitter-type behaviour at large distances. These solutions are also
characterised by a finite GB term and finite, constant components of the
energy-momentum tensor at the far asymptotic regime.

%%%%%%%%%%%%%%%%%%%%%%%%%%%%%%%%%%%%%%%%%%%%%%%%%%%%%%

\subsection{Thermodynamical Analysis}

In this subsection, we calculate the thermodynamical properties of the sought-for 
black-hole solutions, namely their temperature and entropy. The first quantity may be
easily derived by using the following definition \cite{York, GK}
%%%%%%%%%%%%%%%
\beq 
T=\frac{k_h}{2\pi}=\frac{1}{4\pi}\,\left(\frac{1}{\sqrt{|g_{tt} g_{rr}|}}\,
\left|\frac{dg_{tt}}{dr}\right|\right)_{r_h}=\frac{\sqrt{a_1 b_1}}{4\pi}\,,
\label{Temp-def}
\eeq
%%%%%%%%%%%%%%%%
that relates the black-hole temperature $T$ to its surface gravity $k_h$. The above formula
is valid for spherically-symmetric black holes in theories that may contain also higher-derivative
terms such as the GB term. The final expression of the temperature in Eq. (\ref{Temp-def})
is derived by employing the near-horizon asymptotic forms (\ref{A-rh})-(\ref{B-rh}) of the
metric functions.

The entropy of the black hole may be calculated by using the Euclidean approach in which
the entropy is given by the relation \cite{GH}
%%%%%%%%%%%%%%%%%%%
\beq
S_h=\beta\left[\frac{\partial (\beta F)}{\partial \beta} -F\right],
\label{entropy-def}
\eeq
%%%%%%%%%%%%%%%%%%%%
where $F=I_E/\beta$ is the Helmholtz free-energy of the system given in terms of the
Euclidean version of the action $I_E$, and $\beta=1/(k_B T)$. The above formula has been
used in the literature to determine the entropy of the asymptotically-flat coloured GB black
holes \cite{KT} and of the family of novel black-hole solutions found in \cite{ABK} for
different forms of the GB coupling function. However, in the case of a non-asymptotically-flat
behaviour, the above method needs to be modified: in the case of a de-Sitter-type
asymptotic solution, the Euclidean action needs to be integrated only over the causal spacetime
$r_h \leq r \leq r_c$ whereas, for an Anti-de Sitter-type asymptotic solution, the Euclidean action
needs to be regularised \cite{HP, Dutta}, by subtracting the diverging, `pure' AdS-spacetime contribution.

Alternatively, one may employ the Noether current approach developed in \cite{Wald} to calculate
the entropy of a black hole. In this, the Noether current of the theory under diffeomorphisms
is determined, with the Noether charge on the horizon being identified with the entropy of the
black hole. In \cite{Iyer}, the following formula was finally derived for the entropy 
%%%%%%%%%%%
\begin{equation}
S=-2\pi \oint{d^2x \sqrt{h_{(2)}}\left(\frac{\partial \mathcal{L}}{\partial R_{abcd}}\right)_\mathcal{H}\hat{\epsilon}_{ab}\,\hat{\epsilon}_{cd}}\,,
\label{entropy_AdS}
\end{equation}
%%%%%%%%%%%%%%
where $\mathcal{L}$ is the Lagrangian of the theory, $\hat{\epsilon}_{ab}$ the binormal to
the horizon surface $\mathcal{H}$, and $h_{(2)}$ the 2-dimensional projected metric on 
$\mathcal{H}$. The equivalence of the two approaches has been demonstrated in \cite{Dutta},
in particular in the context of theories that contain higher-derivative terms such as the GB term. 
Here, we will use the Noether current approach to calculate the entropy of the black holes
as it leads faster to the desired result. 

To this end, we need to calculate the derivatives of the scalar gravitational quantities, appearing
in the Lagrangian of our theory (\ref{action}), with respect to the Riemann tensor. In Appendix
\ref{variation}, we present a simple way to derive those derivatives. Then, substituting in
Eq. (\ref{entropy_AdS}), we obtain
%%%%%%%%%%%%%%%%%%%%
\begin{align}
S=&-\frac{1}{8}\oint{d^2x \sqrt{h_{(2)}}\bigg\{\frac{1}{2}\left(g^{ac}g^{bd}-g^{bc}g^{ad}\right)+f(\phi)\Big[2R^{abcd}+}\nonumber\\
&-2\left(g^{ac}R^{bd}-g^{bc}R^{ad}-g^{ad}R^{bc}+g^{bd}R^{ac}\right)+R\left(g^{ac}g^{bd}-g^{bc}g^{ad}\right)\Big]\bigg\}_\mathcal{H}\hat{\epsilon}_{ab}\,\hat{\epsilon}_{cd}\,.
\label{entropy_1}
\end{align}
The first term inside the curly brackets of the above expression comes from the variation
of the Einstein-Hilbert term and leads to:
%%%%%%%%%%%%%%%%%%%%
\begin{equation}
S_1=-\frac{1}{16}\oint{d^2x\sqrt{h_{(2)}}\left(\hat{\epsilon}_{ab}\,
\hat{\epsilon}^{\,ab}-\hat{\epsilon}_{ab}\,\hat{\epsilon}^{\,ba}\right)}.
\end{equation}
We recall that $\hat{\epsilon}_{ab}$ is antisymmetric, and, in addition, satisfies 
$\hat{\epsilon}_{ab}\,\hat \epsilon^{\,ab}=-2$. Therefore, we easily obtain the result
%%%%%%%%%%%%%%%
\begin{equation}
S_1=\frac{A_{\mathcal{H}}}{4}. \label{S1}
\end{equation}
%%%%%%%%%%%%%
where $A_{\mathcal{H}}=4\pi r_h^2$ is the horizon surface. The remaining terms in
Eq. (\ref{entropy_1}) are all proportional to the coupling function $f(\phi)$ and follow
from the variation of the GB term. To facilitate the calculation, we notice that, on the
horizon surface, the binormal vector is written as:
$\hat{\epsilon}_{ab}=\sqrt{-g_{00}\,g_{11}}\big|_{\mathcal{H}} \left(\delta^0_a\delta^1_b-\delta^1_a\delta^0_b\right)$.
This means that we may alternatively write:
%%%%%%%%%%%%%%
\beq
\left(\frac{\partial \mathcal{L}}{\partial R_{abcd}}\right)_\mathcal{H}\hat{\epsilon}_{ab}\,\hat{\epsilon}_{cd}=4g_{00}\,g_{11}\big|_{\mathcal{H}}\left(\frac{\partial \mathcal{L}}{\partial R_{0101}}\right)_\mathcal{H}.
\eeq
%%%%%%%%%%%%%%
Therefore, the terms proportional to $f(\phi)$ may be written as
%%%%%%%%%%%%%%%%%%
\bea
S_2&=&-\frac{1}{2}\,f(\phi)\,g_{00}\,g_{11}\big|_{\mathcal{H}}\,\oint d^2x \sqrt{h_{(2)}}\,\bigg[2R^{0101} \nonumber \\[2mm] && \hspace*{3cm}-2\left(g^{00}R^{11}-g^{10}R^{01}-g^{01}R^{10}+g^{11}R^{00}\right)+g^{00}g^{11}R\bigg]_{\mathcal{H}}\,.
\label{S2}
\eea
%%%%%%%%%%%%%%%%%%
To evaluate the above integral, we will employ the near-horizon asymptotic solution
(\ref{A-rh})-(\ref{phi-rh}) for the metric functions and scalar field. The asymptotic values of
all quantities appearing inside the square brackets above are given in Appendix \ref{variation}.
Substituting in Eq. (\ref{S2}), we straightforwardly find 
%%%%%%%%%%%%%%%%
\begin{equation}
S_2=\frac{f(\phi_h)A_{\mathcal{H}}}{r_h^2}= 4\pi f(\phi_h). \label{S2-final}
\end{equation}
%%%%%%%%%%%%%%
Combining the expressions (\ref{S1}) and (\ref{S2-final}), we finally derive the result
%%%%%%%%%%%%%%%
\beq
S_h=\frac{A_h}{4} +4 \pi f(\phi_h)\,.
\label{entropy}
\eeq
%%%%%%%%%%%%%%
The above describes the entropy of a GB black hole arising in the context of the theory
(\ref{action}), with a general coupling function $f(\phi)$ between the scalar field
and the GB term, and a cosmological constant term. We observe that the above expression
matches the one derived in \cite{ABK} in the context of the theory (\ref{action}) but in
the absence of the cosmological constant. This was, in fact, expected on the basis of
the more transparent Noether approach used here: the $\Lambda$ term does not change
the overall topology of the black-hole horizon and it does not depend on the Riemann
tensor; therefore, no modifications are introduced to the functional form of the entropy of
the black hole due to the cosmological constant. However, the presence of $\Lambda$
modifies in a quantitative way the properties of the black hole and therefore the value
of the entropy, and temperature, of the found solutions.

%%%%%%%%%%%%%%%%%%%%%%%%%%%%%%%%%%%%%%%%%%%%%%%%%%%%%%%%%%%%%%%%%%%%%%%%%%%

\section{Numerical Solutions}

In order to construct the complete black-hole solutions in the context of the theory
(\ref{action}), i.e. in the presence of both the GB and the cosmological constant terms,
we need to numerically integrate the system of Eqs. (\ref{A-sys})-(\ref{phi}). The integration
starts at a distance very close to the horizon of the black hole, i.e. at
$r\approx r_h+\mathcal{O}(10^{-5})$ (for simplicity, we set $r_h=1$). The 
metric function $A$ and scalar field $\phi$ in that regime are described by the asymptotic
solutions (\ref{A-rh}) and (\ref{phi-rh}). The input parameter $\phi'_h$
is uniquely determined through Eq. (\ref{solf}) once the coupling function 
$f(\phi) =\alpha \tilde f(\phi)$ is selected and the values of the remaining parameters
of the model near the horizon are chosen. These parameters appear to be $\alpha$,
$\phi_h$ and $\Lambda$. However, the  first two are not independent: since it is their
combination $\alpha \tilde f(\phi_h)$ that determines the strength of the coupling between
the GB term and the scalar field, a change in the value of one of them may be absorbed in 
a corresponding change to the value of the other; as a result, we may fix $\alpha$
and vary only $\phi_h$. The values of $\phi_h$ and $\Lambda$ also cannot be
totally uncorrelated as they both appear in the expression of $C$, Eq. (\ref{C-def}),
that must always be positive; therefore, once the value of the first is chosen,
there is an allowed range of values for the second one for which black-hole solutions
arise. This range of values are determined by the inequalities $\dot f_h^2 \leq \dot f_-^2$
and $\dot f_h^2 \geq \dot f_+^2$ according to Eq. (\ref{C-newdef}), and lead in
principle to two distinct branches of solutions. In fact, removing the square, four
branches emerge depending on the sign of $\dot f_h$. However, in what follows
we will assume that $\dot f_h>0$, and thus study the two regimes $\dot f_h \leq \dot f_-$
and $\dot f_h \geq \dot f_+$; similar results emerge if one assumes instead that
$\dot f_h <0$. 

Before starting our quest for black holes with an (Anti)-de Sitter asymptotic behaviour
at large distances, we first consider the case with $\Lambda=0$ where upon we
successfully reproduce the families of asymptotically-flat back holes derived in 
\cite{ABK}. Then, we select non-vanishing values of $\Lambda$ and look for
novel black-hole solutions. We will start with the case of a negative cosmological
constant ($\Lambda<0$) in the next subsection and consider the case of a positive
cosmological constant ($\Lambda>0$) in the following one.

\subsection{Anti-de Sitter Gauss-Bonnet Black Holes}

As mentioned above, the integration starts from the near-horizon regime with the
asymptotic solutions (\ref{A-rh}) and (\ref{phi-rh}), and it proceeds towards large
values of the radial coordinate until the form of the derived solution for the metric
resembles, for $\Lambda<0$, the asymptotic solution (\ref{alfar1})-(\ref{bfar1})
describing an Anti-de Sitter-type background. The arbitrary coefficient $a_1$,
that does not appear in the field equations, may be fixed by demanding that, at
very large distances, the metric functions satisfy the constraint $e^A \simeq e^{-B}$. 
We have considered a large number of
forms for the coupling function $f(\phi)$, and, as we will now demonstrate, we have
managed to produce a family of regular black-hole solutions with an Anti-de Sitter
asymptotic behaviour, for every choice of $f(\phi)$.

%%%%%%%%%%%%%%%%%%%%%
\begin{figure}[t]
\begin{center}
\mbox{\hspace*{-0.7cm} \includegraphics[width = 0.46 \textwidth] {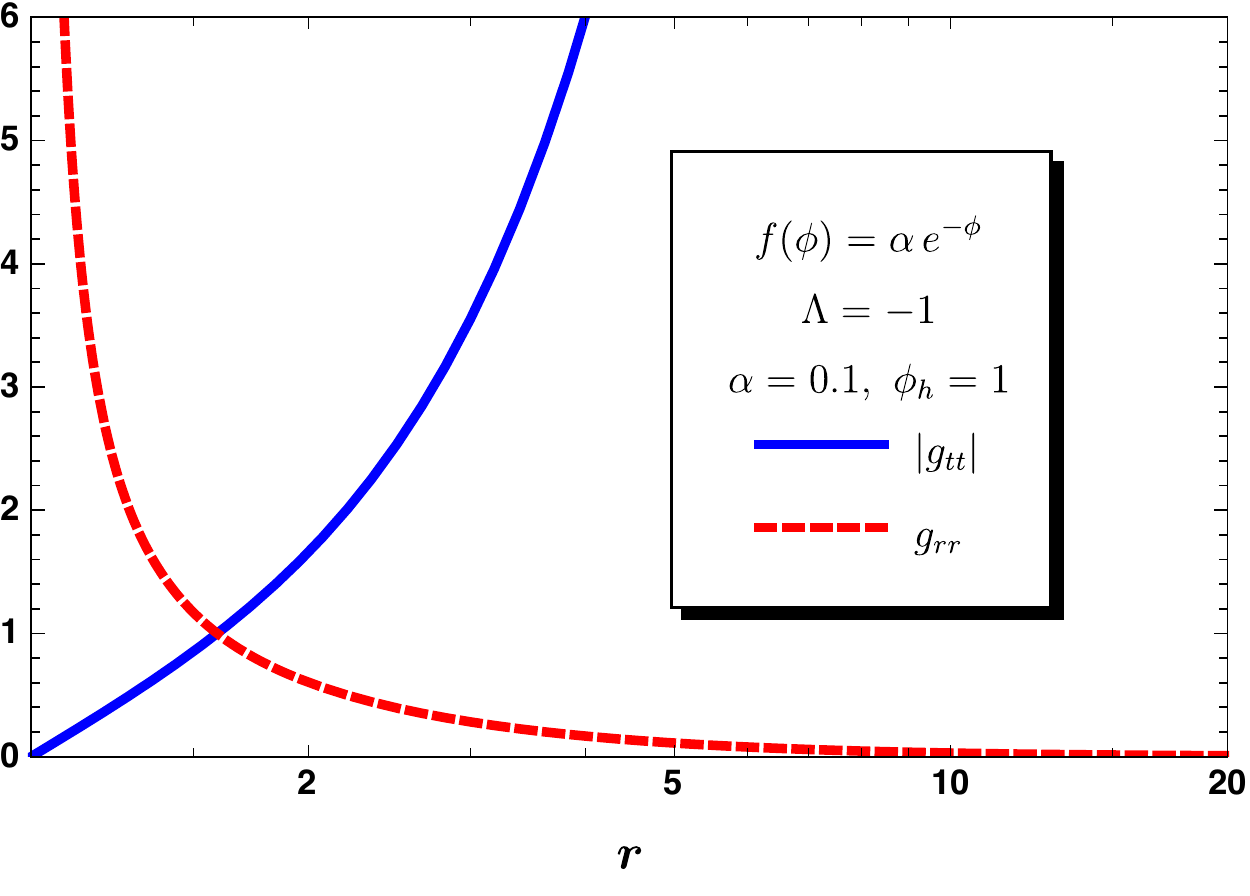}}
\hspace*{-0.0cm} {\includegraphics[width = 0.505 \textwidth]
{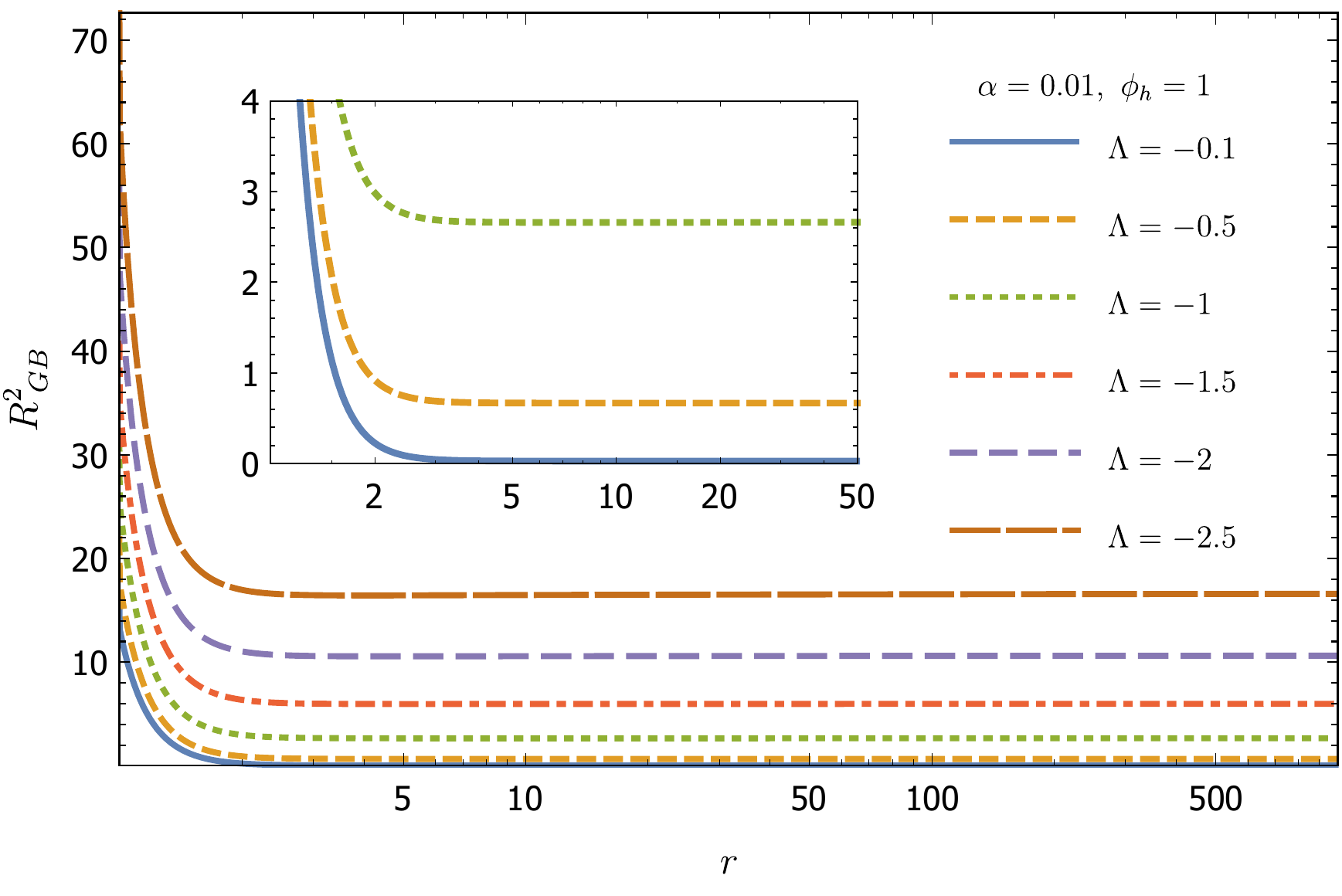}}
    \caption{The metric components $|g_{tt}|$ and $g_{rr}$ (left plot), and the Gauss-Bonnet
term $R^2_{GB}$ (right plot) in terms of the radial coordinate $r$, for $f(\phi)=\alpha e^{-\phi}$.}
   \label{Metric_GB}
  \end{center}
\end{figure}
%%%%%%%%%%%%%%%%%

We will first discuss the case of an exponential coupling function, $f(\phi)=\alpha e^{-\phi}$. The
solutions for the metric functions $e^{A(r)}$ and $e^{B(r)}$ are depicted in the left plot of
Fig. \ref{Metric_GB}. We may easily see that the near-horizon behaviour, with $e^{A(r)}$
vanishing and $e^{B(r)}$ diverging, is eventually replaced by an Anti-de Sitter regime with
the exactly opposite behaviour of the metric functions at large distances. The solution presented
corresponds to the particular values $\Lambda=-1$ (in units of $r_h^{-2}$), $\alpha=0.1$
and $\phi_h=1$, however,
we obtain the same qualitative behaviour for every other set of parameters satisfying the
constraint~\footnote{Here, we do not present black-hole solutions that satisfy the alternative
choice $\dot f_h \geq \dot f_+$ since this leads to solutions plagued by numerical instabilities,
that prevent us from deducing their physical properties in a robust way. The same ill-defined
behaviour of this second branch of solutions with very small horizon radii was also found in
\cite{Hartmann}.} $\dot f_h \leq \dot f_-$, that follows from Eq. (\ref{C-def}). The
spacetime is regular in the whole radial regime, and this is reflected in the form of
the scalar-invariant Gauss-Bonnet term: this is presented in the right plot of
Fig. \ref{Metric_GB}, for $\alpha=0.01$, $\phi_h=1$ and for a variety of values
of the cosmological constant. We observe that the GB term acquires its maximum
value near the horizon regime, where the curvature of spacetime is larger, and
reduces to a smaller, constant asymptotic value in the far-field regime. This asymptotic
value is, as expected, proportional to the cosmological constant as this quantity determines
the curvature of spacetime at large distances.  

\begin{figure}[t]
\minipage{0.48\textwidth}
\hspace*{-0.25cm}  \includegraphics[width=\linewidth]{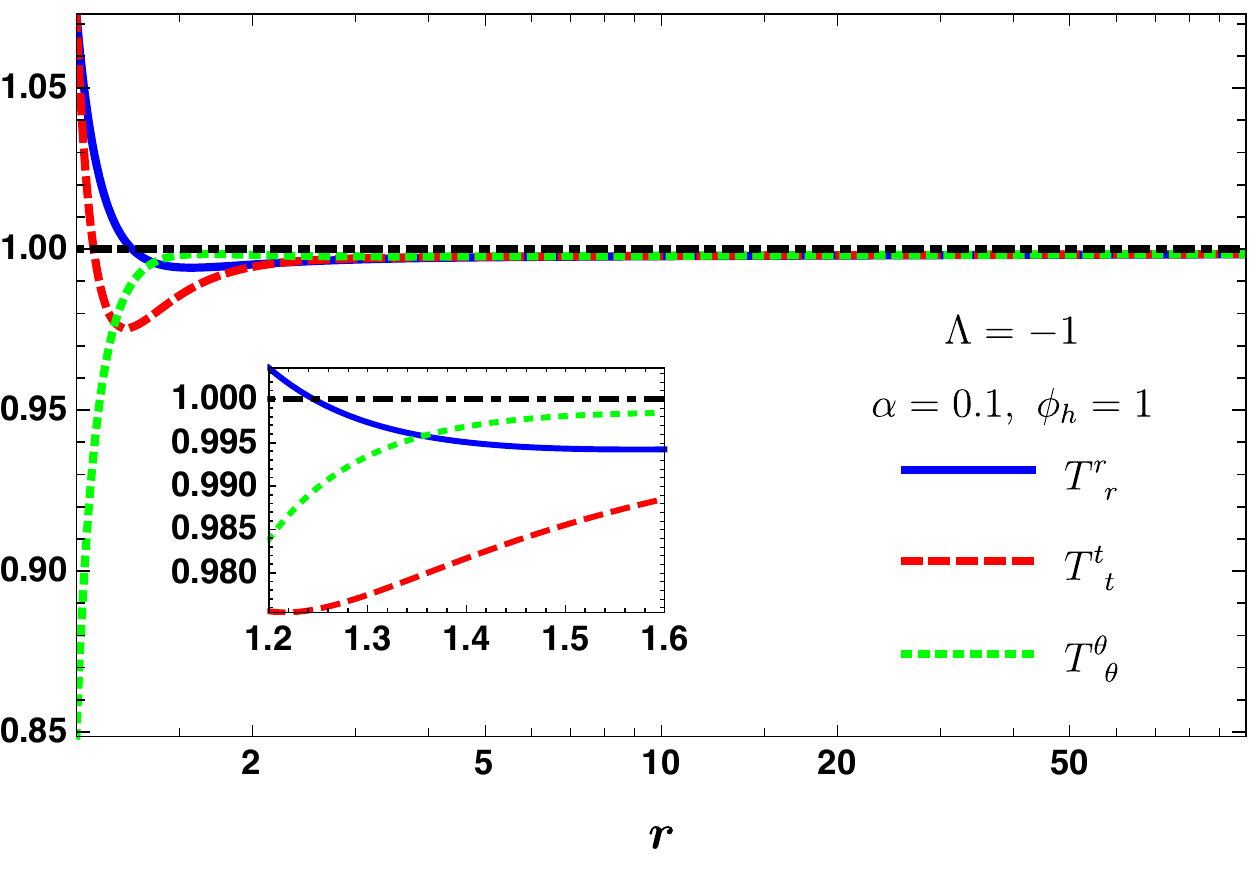}
\endminipage\hfill
\minipage{0.50\textwidth}
\hspace*{-0.4cm}  \includegraphics[width=\linewidth]{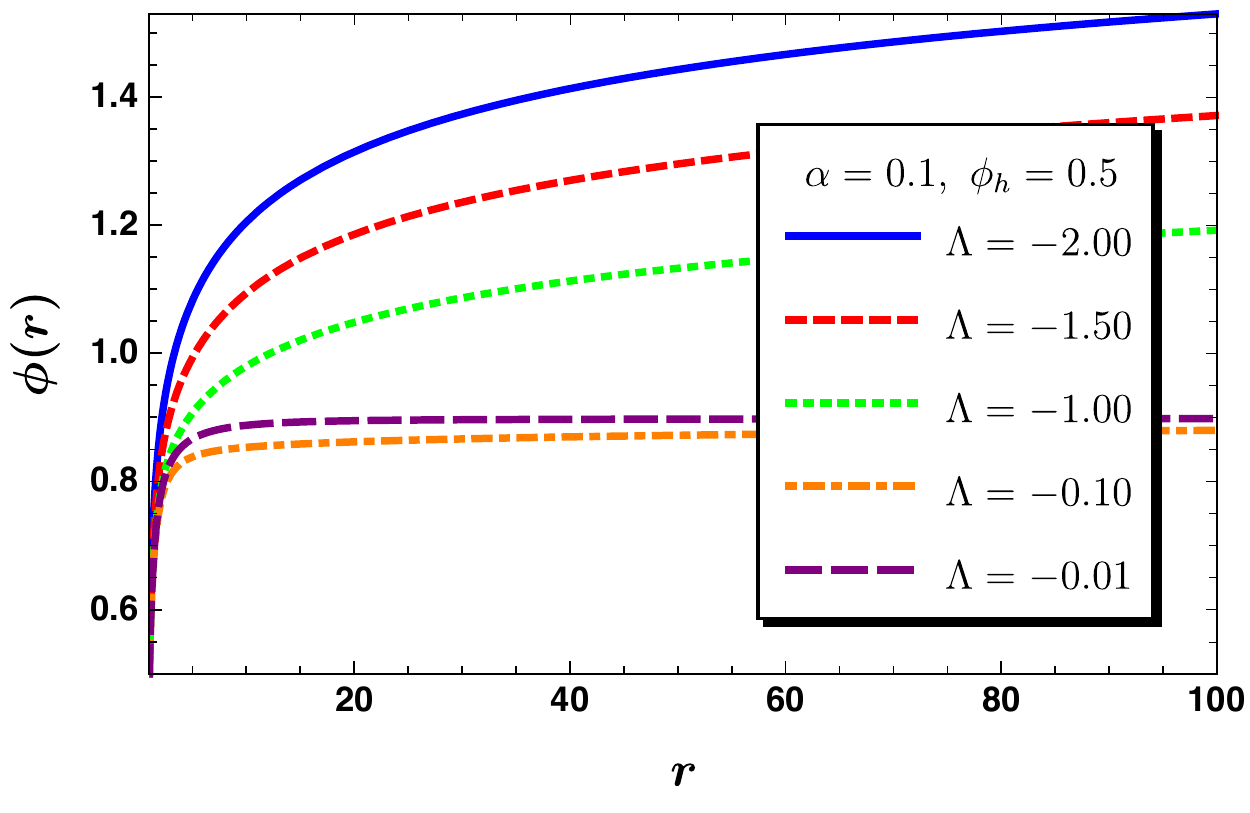}
\endminipage\hfill
 \caption{The energy-momentum tensor $T_{\mu\nu}$ (left plot), and scalar field $\phi$
(right plot) in terms of the radial coordinate $r$, for $f(\phi)=\alpha e^{-\phi}$.}
\label{exp-phi-Tmn}
\end{figure}

Although in Section 2.2.2, we could not find the analytic form of the scalar field at large
distances from the black-hole horizon for different forms of the coupling function $f(\phi)$,
our numerical results ensure that its behaviour is such that the effect of the scalar field
at the far-field regime is negligible, and it is only the cosmological term that determines
the components of the energy-momentum tensor. In the left plot of Fig. \ref{exp-phi-Tmn},
we display all three components of $T^{\mu}_{\;\,\nu}$ over the whole radial regime, for
the indicative solution $\Lambda=-1$, $\alpha=0.1$ and $\phi_h=1$.
Far away from the black-hole horizon, all components reduce to $-\Lambda$, in
accordance with Eqs. (\ref{Ttt})-(\ref{Tthth}), with the effect of both the scalar field and
the GB term being there negligible. Near the horizon, and according to the asymptotic behaviour
given by Eqs. (\ref{Ttt_rh})-(\ref{Tthth_rh}), we always have $T^r_{\;\,r}\approx T^t_{\;\,t}$,
since, at $r \simeq r_h$, $A' \simeq -B'$; also, the $T^\theta_{\;\,\theta}$ component
always has the opposite sign to that of $T^r_{\;\,r}$ since $A'' \simeq -A'^2$. This
qualitative behaviour of $T^{\mu}_{\;\,\nu}$ remains the same for all forms of the coupling
function we have studied and for all solutions found, therefore we refrain from giving
additional plots of this quantity for the other classes of solutions found.

%%%%%%%%%%%%%%%%%%%%%%%%%%
%%%%%%%%%%%%%%%%%%%%%%%%%%

From the results depicted in the left plot of Fig. \ref{exp-phi-Tmn}, we see that, near the
black-hole horizon, we always have $T^r_{\;\,r}\approx T^t_{\;\,t}>0$. Comparing this
behaviour with the asymptotic forms (\ref{Ttt_rh})-(\ref{Tthth_rh}), we deduce that, close
to the black-hole horizon where $A'>0$, we must have $(\phi' \dot f)_h <0$. 
In the case of vanishing cosmological constant, the negative value of this quantity was of
paramount importance for the evasion of the no-hair theorem \cite{Bekenstein} and the
emergence of novel, asymptotically-flat black-hole solutions \cite{ABK}. We observe that also in
the context of the present analysis with $\Lambda \neq 0$, this quantity turns out to be again
negative, and to lead once again to novel black-hole solutions. Coming back to our assumption
of a decreasing exponential coupling function and upon choosing to consider $\alpha>0$,
the constraint
$(\phi' \dot f)_h <0$ means that $\phi'_h>0$ independently of the value of $\phi_h$. In
the right plot of Fig. \ref{exp-phi-Tmn}, we display the solution for the scalar field in
terms of the radial coordinate, for the indicative values of $\alpha=0.1$, $\phi_h=0.5$ and
for different values of the cosmological constant. The scalar field satisfies indeed the 
constraint $\phi'_h>0$ and increases away from the black-hole horizon\footnote{A
complementary family of  solutions arises if we choose $\alpha<0$, with the scalar
profile now satisfying the constraint $\phi'_h<0$ and decreasing away from the black-hole
horizon.}. At large distances, we observe that, for small values of the cosmological
constant, $\phi(r)$ assumes a constant value; this is the behaviour found for asymptotically-flat
solutions \cite{ABK} that the solutions with small $\Lambda$ are bound to match.
For increasingly larger values of $\Lambda$ though, the profile of the scalar field deviates
significantly from the series expansion in powers of $(1/r)$ thus allowing for a $r$-dependent
$\phi$ even at infinity -- in the perturbative limit, as we showed in the previous section,
this dependence is given by the form $\phi(r) \simeq d_1 \ln r$.

%%%%%%%%%%%%%%%%%%%%%%%%%%%%%%%%%%%%%%%%%%

\begin{figure}[t]
\minipage{0.50\textwidth}
\hspace*{-0.4cm}   \includegraphics[width=\linewidth]{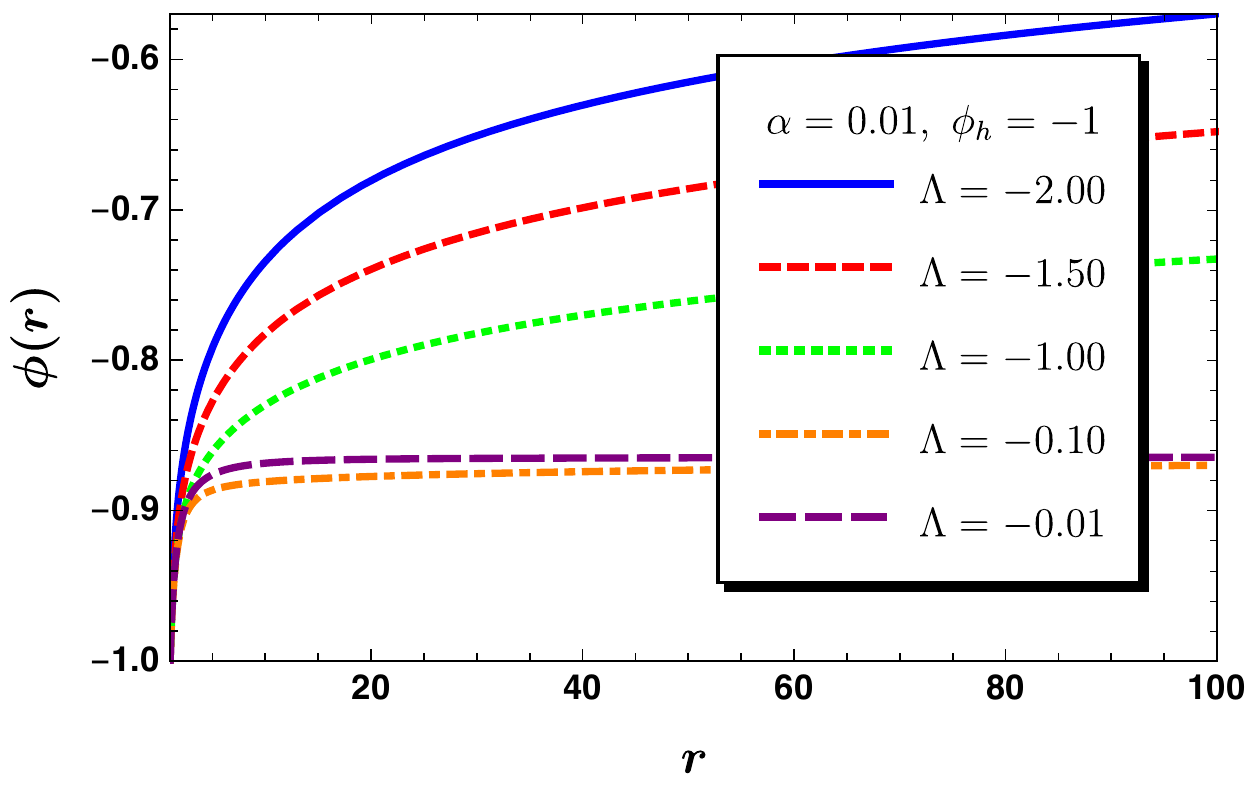}
\endminipage\hfill
\minipage{0.50\textwidth}
\hspace*{-0.4cm}   \includegraphics[width=\linewidth]{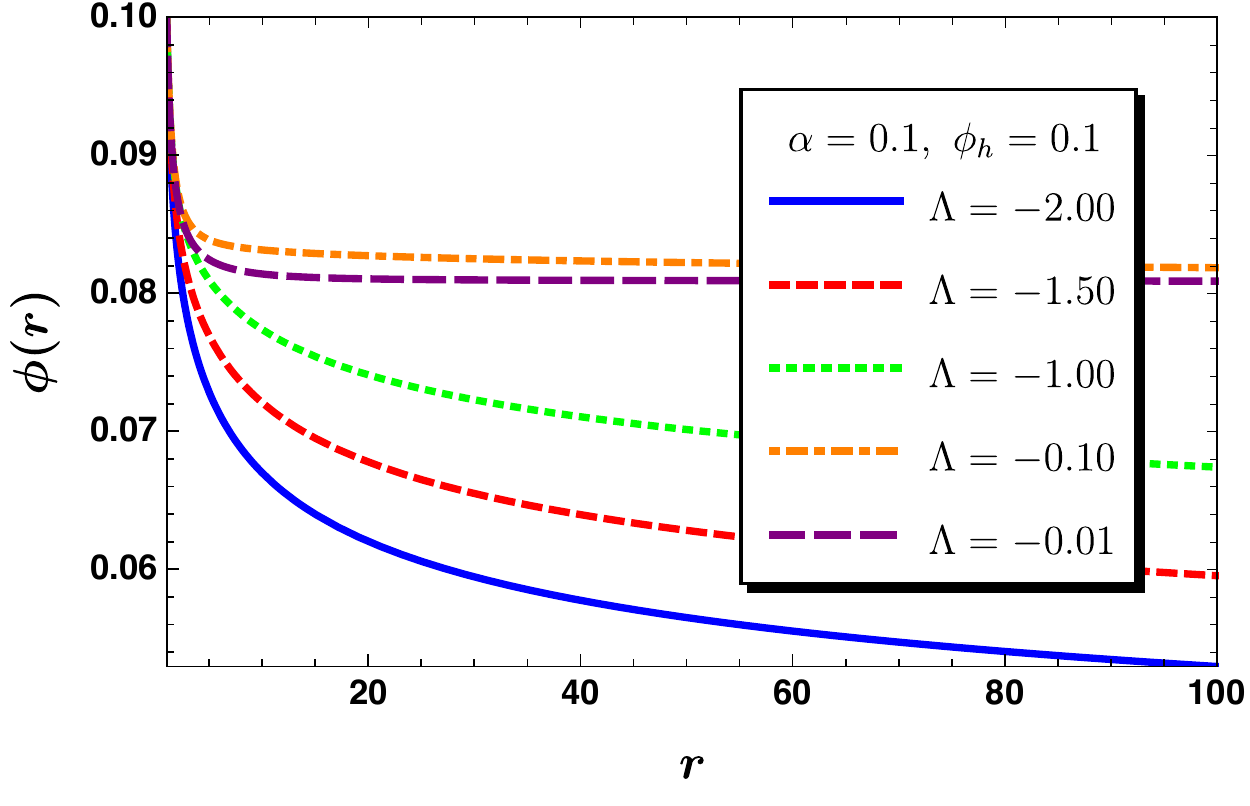}
\endminipage\hfill
 \caption{The scalar field $\phi$  in terms of the radial coordinate $r$, for
$f(\phi)=\alpha\phi^2$ (left plot) and $f(\phi)=\alpha\phi^3$ (right plot).}
\label{fig_quad_qub}
\end{figure}

We will now consider the case of an even polynomial coupling function of the form
$f(\phi)=\alpha \phi^{2n}$ with $n \geq 1$. The behaviour of the solution for the metric
functions matches the
one depicted\,\footnote{Let us mention at this point that, for extremely large values of either
the coupling constant $\alpha$ or the cosmological constant $\Lambda$, that are nevertheless
allowed by the constraint (\ref{C-def}), solutions that have their metric behaviour deviating
from the AdS-type form (\ref{alfar1})-(\ref{bfar1}) were found; according to the obtained
behaviour, both metric functions seem to depend logarithmically on the radial coordinate
instead of polynomially. As the physical interpretation of these solutions is not yet clear,
we omit these solutions from the remaining of our analysis.} in the left plot of Fig.
\ref{Metric_GB}. The same is true for the behaviour of the GB term and the energy-momentum tensor,
whose profiles are similar to the ones displayed in Figs. \ref{Metric_GB} (right plot) and
\ref{exp-phi-Tmn} (left plot), respectively. The positive-definite value of $T^r_{\,\,r}$
near the black-hole horizon implies again that, there, we should have $(\dot{f}\phi')_h<0$,
or equivalently $\phi_h\phi'_h<0$,  for $\alpha>0$. Indeed, two classes of solutions arise
in this case: for positive values of $\phi_h$, we obtain solutions for the scalar field that
decrease away from the black-hole horizon, while for $\phi_h<0$, solutions that increase with
the radial coordinate are found. In Fig. \ref{fig_quad_qub} (left plot), we present a family of
solutions for the case of the quadratic coupling function (i.e. $n=1$), for $\phi_h=-1$ and
$\alpha=0.01$, arising for different values of $\Lambda$ -- since $\phi_h<0$, the scalar
field exhibits an increasing behavior as expected.

%%%%%%%%%%%%%%%%%%%%%%
%%%%%%%%%%%%%%%%%%%%%%

%%%%%%%%%%%%%%%%%%%%%%%%%%%%%%%%%%%%%

Let us examine next the case of an odd polynomial coupling function, $f(\phi)=\alpha\phi^{2n+1}$
with $n \geq 0$. The behaviour of the metric functions, GB term and energy-momentum tensor have
the expected behaviour for an asymptotically AdS background, as in the previous cases. The solutions
for the scalar field near the black-hole horizon are found to satisfy the constraint
$\alpha (\phi^{2n}\phi')_h<0$ or simply $\phi_h'<0$, when $\alpha>0$. As this holds independently of
the value of $\phi_h$, all solutions for the scalar field are expected to decrease away from the
black-hole horizon. Indeed, this is the profile depicted in the right plot of Fig. \ref{fig_quad_qub}
where a family of solutions for the indicative case of a qubic coupling function (i.e. $n=1$) is
presented for $\alpha=0.1$, $\phi_h=0.1$ and various values of $\Lambda$. 

%%%%%%%%%%%%%%%%%%%%%%%%%%%%%%%%%%%%%%%%%%%%%%%%%%%%%%

\begin{figure}[t!]
\minipage{0.50\textwidth}
\hspace*{-0.4cm}   \includegraphics[width=\linewidth]{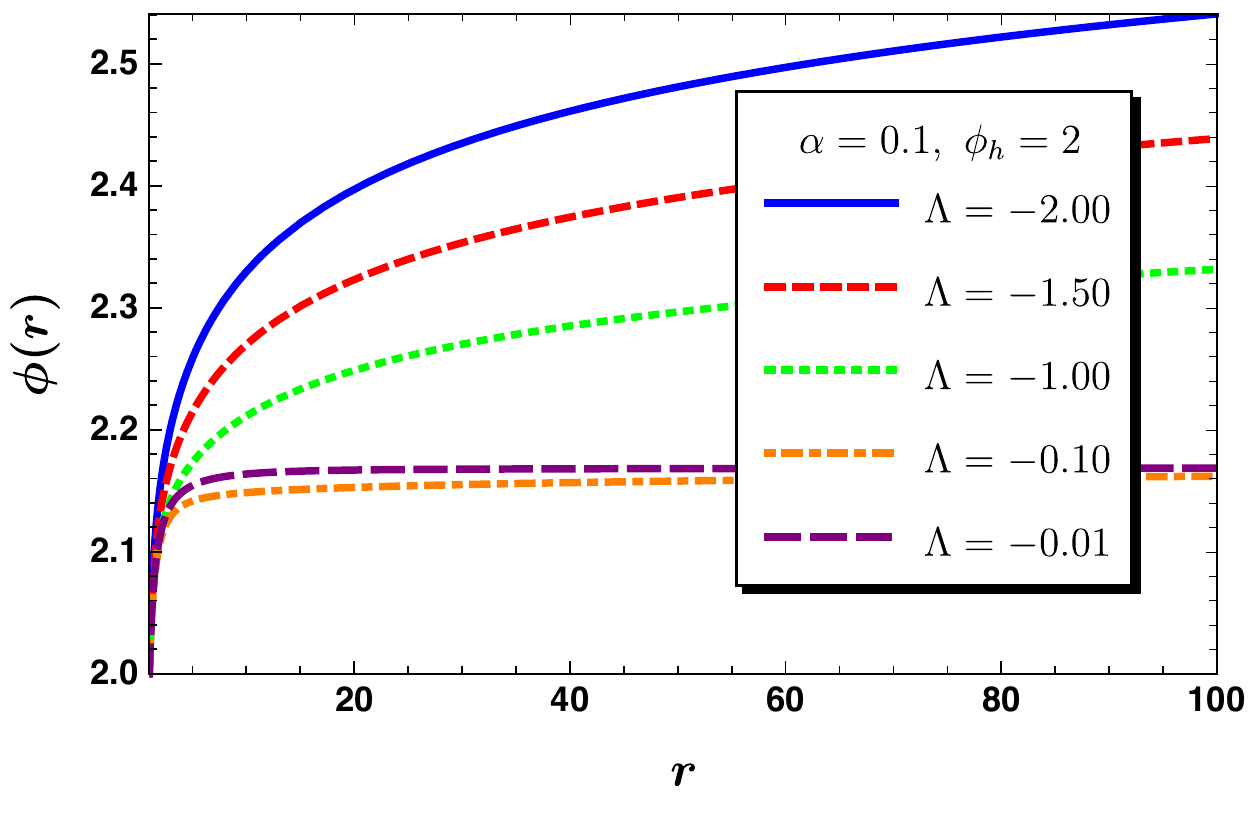}
\endminipage\hfill
\minipage{0.50\textwidth}
\hspace*{-0.4cm}   \includegraphics[width=\linewidth]{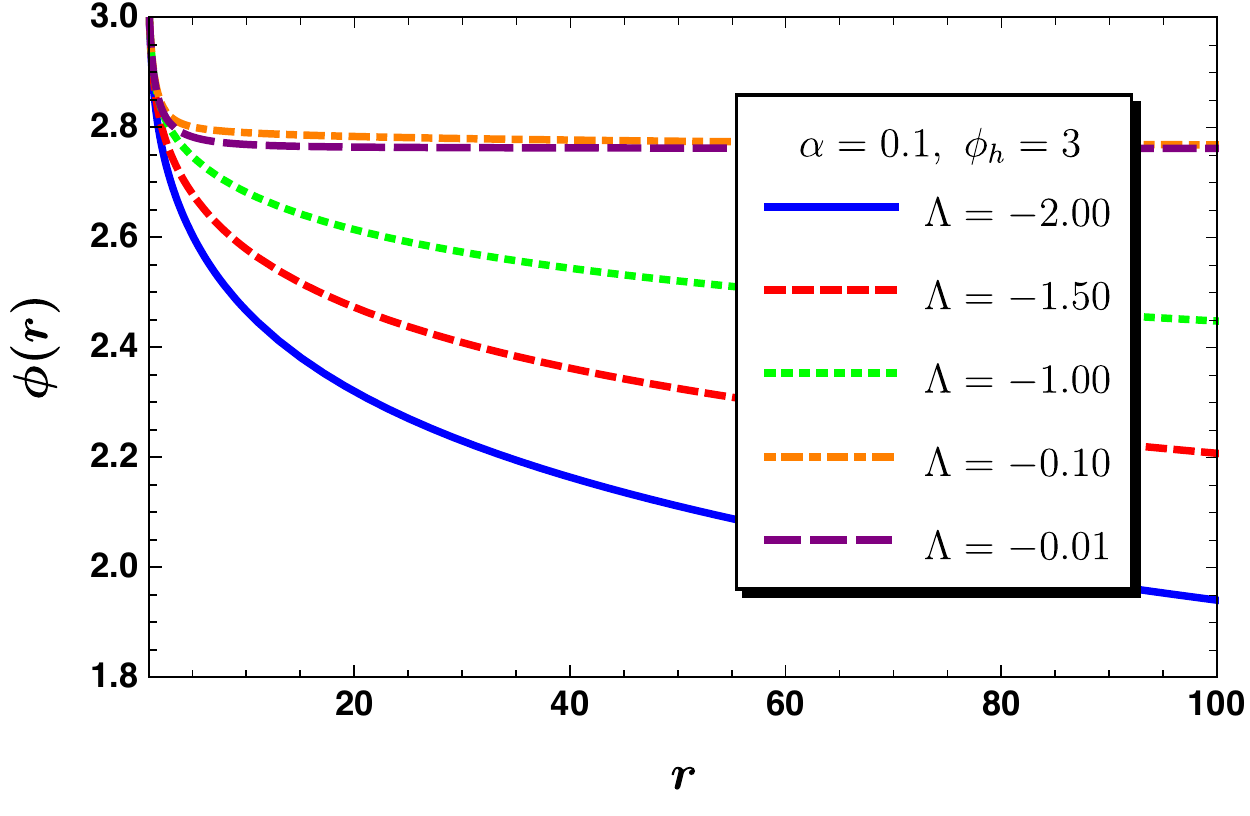}
\endminipage\hfill
\caption{The scalar field $\phi$  in terms of the radial coordinate $r$, for
$f(\phi)=\alpha/\phi$ (left plot) and $f(\phi)=\alpha\ln\phi$ (right plot).}
\label{fig_inv_log}
\end{figure}

The case of an inverse polynomial coupling function, $f(\phi)=\alpha\phi^{-k}$, with $k$ either an
even or odd positive integer, was also considered. For odd $k$, i.e. $k=2n+1$, the positivity
of $T^r_{\,\,r}$ near the black-hole horizon demands again that $(\dot{f}\phi')_h<0$, or that $-\alpha/\phi^{2n+2}\phi'<0$. For $\alpha>0$, the solution for the scalar field should thus
always satisfy $\phi'_h>0$, regardless of our choices for $\phi_h$ or $\Lambda$. As an indicative
example, in the left plot of Fig. \ref{fig_inv_log}, we present the case of $f(\phi)=\alpha/\phi$
with a family of solutions arising for $\alpha=0.1$ and $\phi_h=2$. The solutions for the
scalar field clearly satisfy the expected behaviour by decreasing away from the black-hole horizon.
On the other hand, for even $k$, i.e. $k=2n$, the aforementioned constraint now demands that
$\phi_h\,\phi'_h<0$. As in the case of the odd polynomial coupling function, two subclasses of solutions
arise: for $\phi_h>0$, solutions emerge with $\phi'_h<0$ whereas, for $\phi_h<0$, we find solutions 
with $\phi'_h>0$. The profiles of the solutions in this case are similar to the ones found before,
with $\phi$ approaching, at large distances, an almost constant value for small $\Lambda$ 
but adopting a more dynamical behaviour as the cosmological constant gradually takes on
larger values. 

%%%%%%%%%%%%%%%%%%%%%%
%%%%%%%%%%

%%%%%%%%%%%%%%%%%%%%%%%%%%%%%%%%%%%%%%%%%%%%%%%%%%%%%%%%%%%%%%%%%%%%%

As a final example of another form of the coupling function between the scalar field and the
GB term, let us consider the case of a logarithmic coupling function, $f(\phi)=\alpha \ln\phi$.
Here, the condition near the horizon of the black hole gives $\alpha \phi'/\phi<0$, therefore,
for $\alpha>0$, we must have $\phi'_h\phi_h<0$; for $\phi_h>0$, this translates to a
decreasing profile for the scalar field near the black-hole horizon. In the right plot of Fig.
\ref{fig_inv_log}, we present a family of solutions arising for a logarithmic coupling function
for fixed $\alpha=0.01$ and $\phi_h=1$, while varying the cosmological constant $\Lambda$.
The profiles of the scalar field agree once again with the one dictated by the near-horizon
constraint, and they all decrease in that regime. As in the previous cases, the metric
functions approach asymptotically an Anti-de Sitter background, the scalar-invariant GB term
remains everywhere regular, and the same is true for all components of the energy-momentum
tensor that asymptotically approach the value $-\Lambda$.

%%%%%%%%%%%%%%%%%%%%%
\begin{figure}[t!]
\begin{center}
\minipage{0.49\textwidth}
  \includegraphics[width=\linewidth]{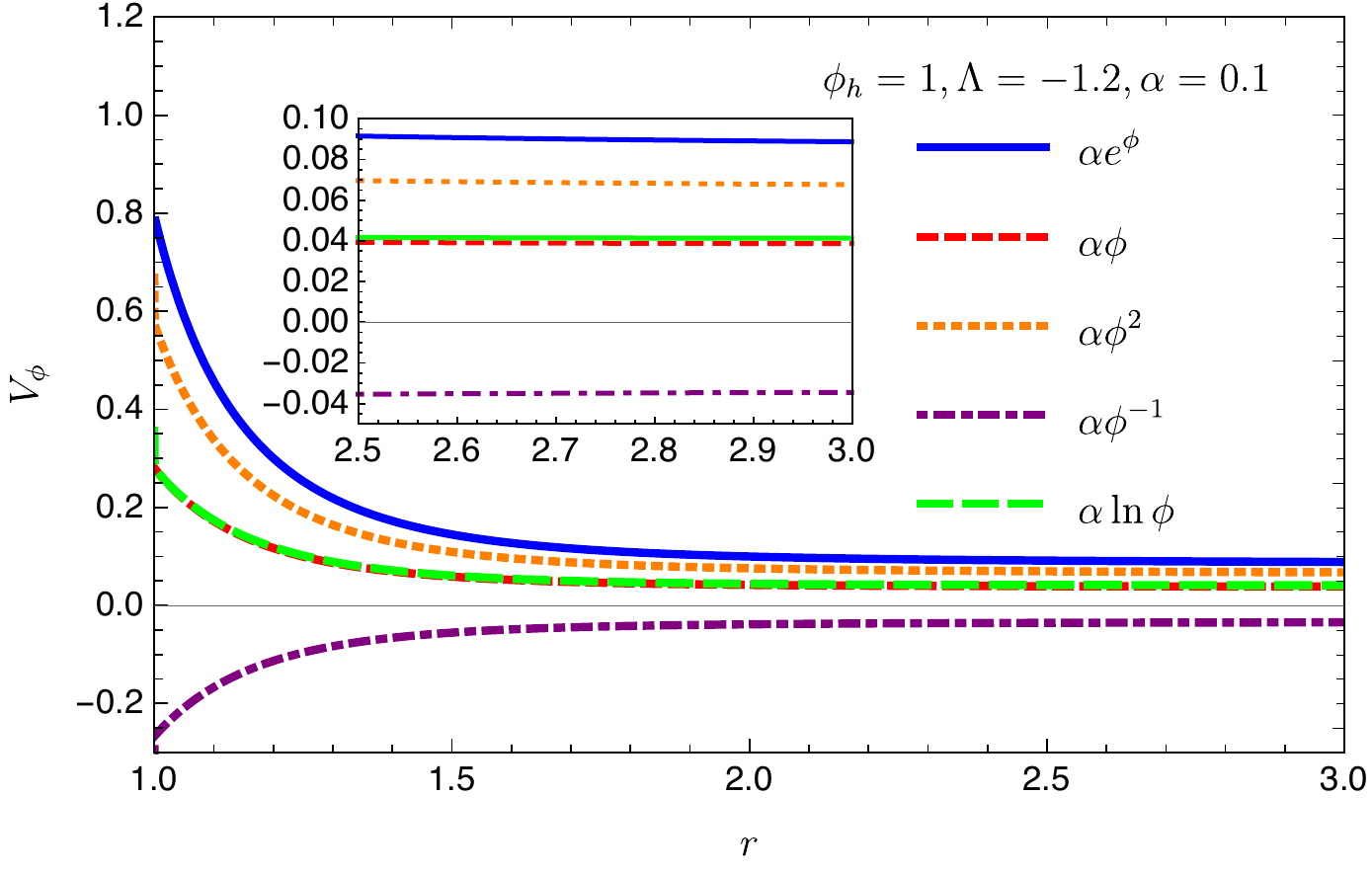}
\endminipage\hfill \hspace*{-0.4cm}
\minipage{0.49\textwidth}
  \includegraphics[width=\linewidth]{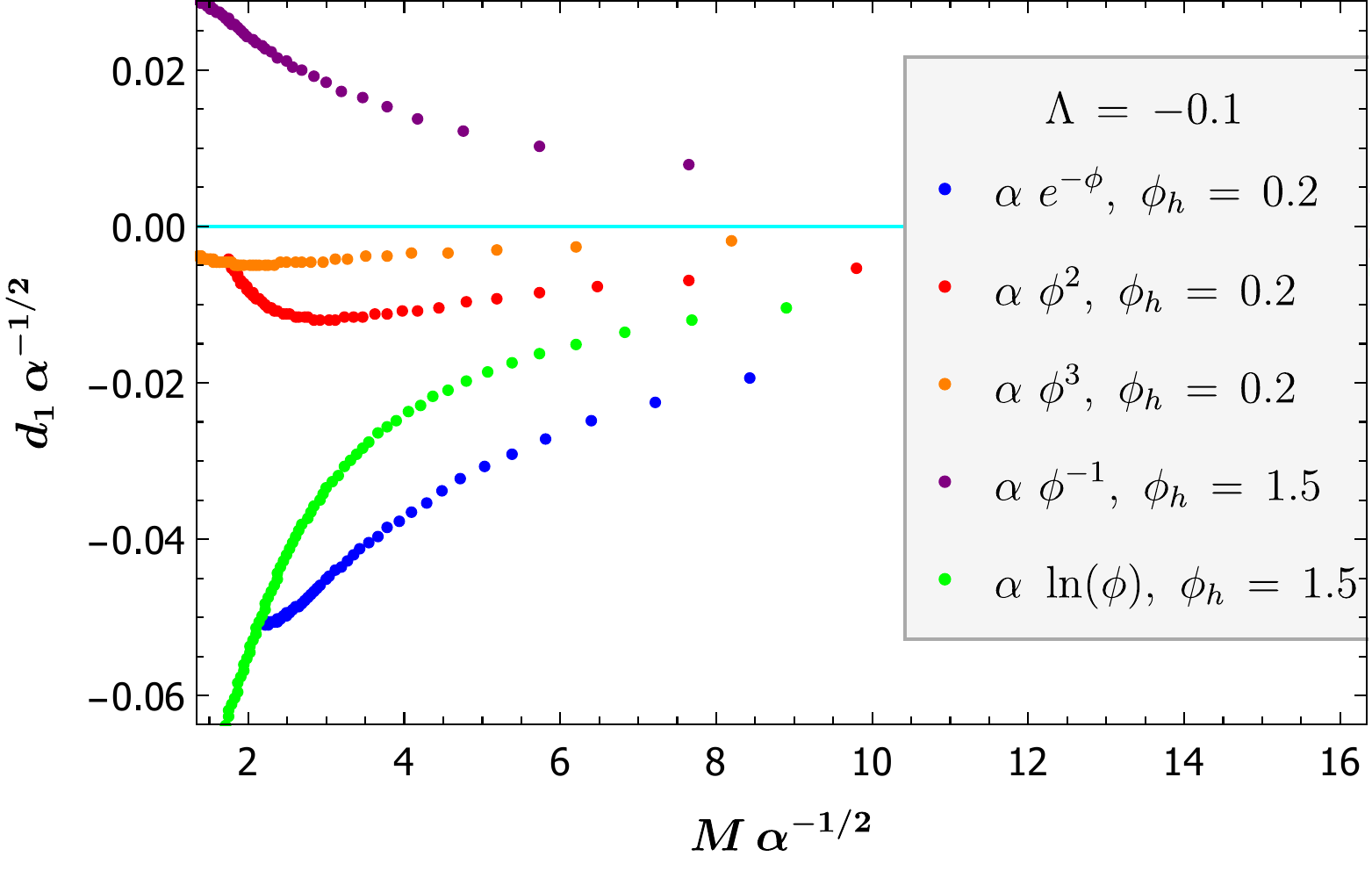}
\endminipage\hfill
    \caption{The effective potential $V_\phi$ of the scalar field, in terms of the
radial coordinate (left plot), and the coefficient $d_1$ (right plot)
in terms of the mass $M$, for various forms of $f(\phi)$.}
   \label{V-d1}
  \end{center}
\end{figure}
%%%%%%%%%%%%%%%%%
%%%%%%%%%%%%%%%%%%%%%%%%%%%%%%%%%%%%%%%%%%%%%%%

It is of particular interest to study also the behaviour of the effective potential of the scalar
field, a role that in our theory is played by the GB term together with the coupling function,
i.e. $V_\phi \equiv \dot f(\phi)\,R^2_{GB}$. In the left plot of Fig. \ref{V-d1}, we present a combined graph
that displays its profile in terms of the radial coordinate, for a variety of forms of the
coupling function $f(\phi)$. As expected, the potential $V_\phi$ takes on its maximum
value always near the horizon of the black hole, where the GB term is also maximized and 
thus sources the non-trivial form of the scalar field. On the other hand, as we move
towards larger distances, $V_\phi$ reduces to an asymptotic constant value. Although
this asymptotic value clearly depends on the choice of the coupling function, its common
behaviour allows us to comment on the asymptotic behaviour of the scalar field at
large distances. Substituting a constant value $V_\infty$ in the place of $V_\phi$ in
the scalar-field equation (\ref{phi-eq}), we arrive at the intermediate result
%%%%%%%%%%%%%%%
\beq
\partial_r \left[e^{(A-B)/2} r^2 \phi' \right] = -e^{(A+B)/2} r^2\,V_\infty\,.
\eeq
%%%%%%%%%%%%%
Then, employing the asymptotic forms of the metric functions at large distances
(\ref{alfar1})-(\ref{bfar1}), the above may be easily integrated with respect to the radial
coordinate to yield a form for the scalar field identical to the one given in 
Eq. (\ref{phi-far-Anti}). We may thus conclude that the logarithmic
form of the scalar field may adequately describe its far-field behaviour even beyond
the perturbative limit of very small $\alpha$. 

We now proceed to discuss the physical characteristics of the derived solutions. Due to
the large number of solutions found, we will present, as for $V_\phi$, combined graphs
for different forms of the coupling function $f(\phi)$. Starting with the scalar field, we
notice that no conserved quantity, such as a scalar charge, may be associated with
the solution at large distances in the case of asymptotically Anti-de Sitter black holes:
the absence of an ${\cal O}(1/r)$ term in the far-field expression (\ref{phi-far-Anti})
of the scalar field, that would signify the existence of a long-range interaction term,
excludes the emergence of such a quantity, even of secondary nature. One could attempt
instead to plot the dependence of the coefficient $d_1$, as a quantity that predominantly
determines the rate of change of the scalar field at the far field, in terms of the mass
of the black hole. This is displayed in the right plot of Fig. \ref{V-d1} for the indicative
value $\Lambda=-0.1$ of the cosmological constant.
We see that, for small values of the mass $M$, this coefficient takes in general a non-zero
value, which amounts to having a non-constant value of the scalar field at the far-field
regime. As the mass of the black hole increases though, this coefficient asymptotically
approaches a zero value. Therefore, the rate of change of the scalar field at infinity
for massive GB black holes becomes negligible and the scalar field tends to a constant.
This is the `Schwarzschild-AdS regime', where the GB term decouples from the theory and
the scalar-hair disappears - the same behaviour was observed also in the case of
asymptotically-flat GB black holes \cite{ABK} where, in the limit of large mass, all
of our solutions merged with the Schwarzschild ones.

\begin{figure}[t!]
\begin{center}
\minipage{0.49\textwidth} \hspace*{-0.2cm}
  \includegraphics[width=\linewidth]{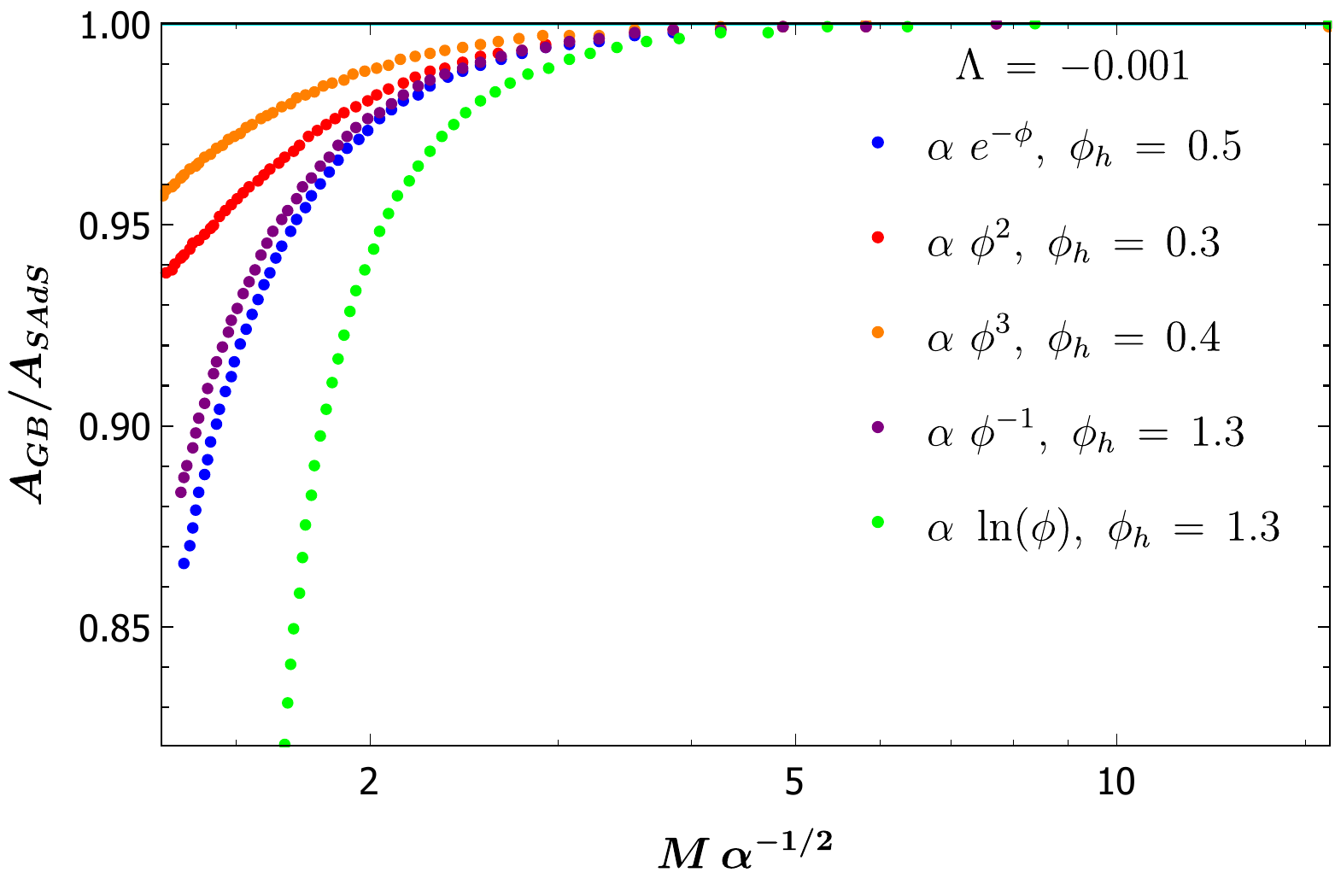}
\endminipage\hfill \hspace*{-1.8cm}
\minipage{0.49\textwidth}
  \includegraphics[width=\linewidth]{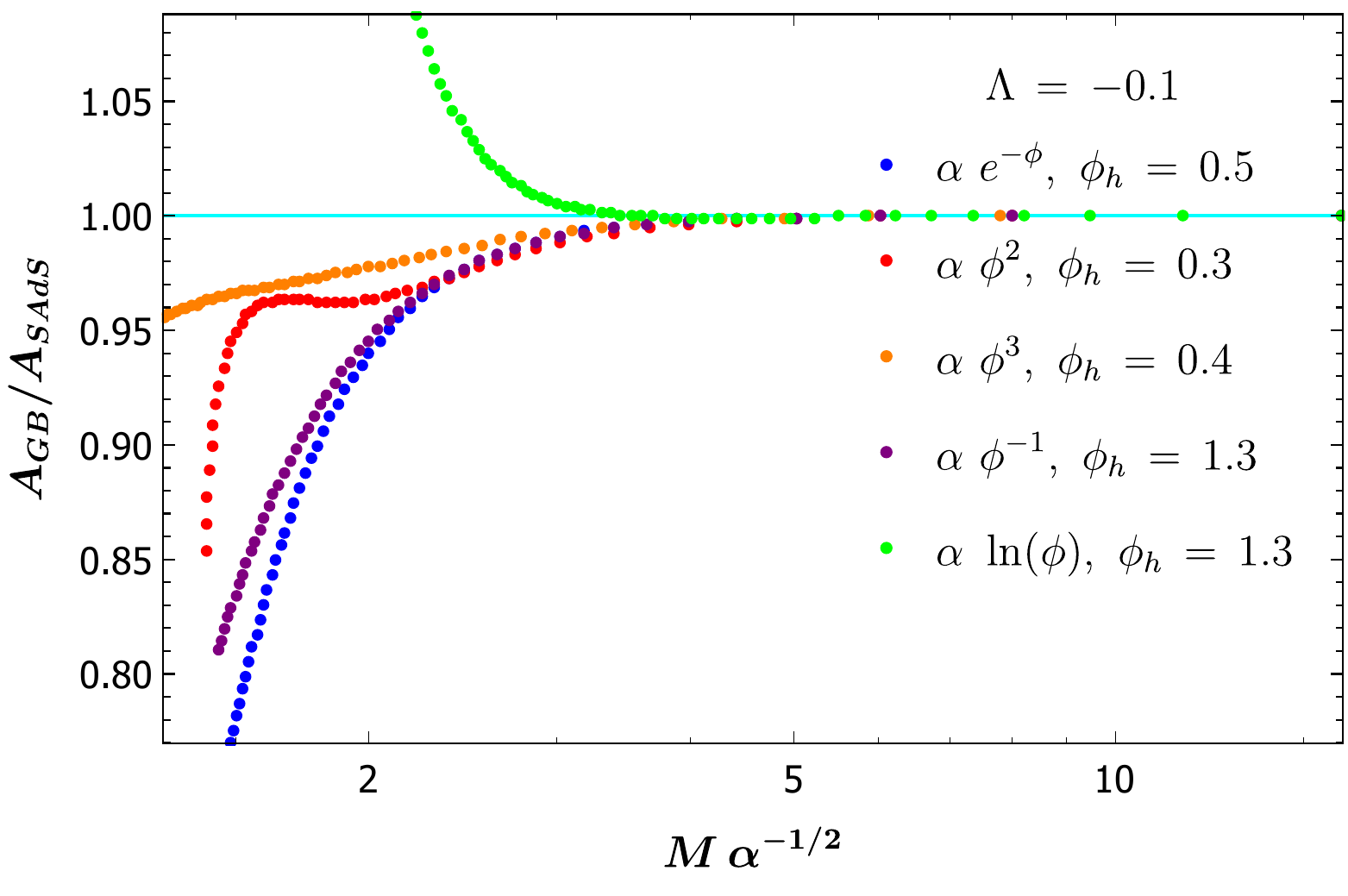}
\endminipage\hfill
    \caption{The area ratio $A_{GB}/A_{SAdS}$ of our solutions as a function of the
mass $M$ of the black hole, for various forms of $f(\phi)$, and for $\Lambda=-0.001$
(left plot) and $\Lambda=-0.1$ (right plot).}
   \label{Area}
  \end{center}
\end{figure}

We present next the ratio of the horizon area of our solutions compared to the
horizon area of the SAdS one with the same mass, for the indicative values of the
negative cosmological constant $\Lambda=-0.001$ and $\Lambda=-0.1$ in the
two plots of Fig. \ref{Area}. These plots provide further evidence for the merging of our
GB black-hole solutions with the SAdS solution in the limit of large mass. The left
plot of Fig. \ref{Area} reveals that, for small cosmological constant, all our GB solutions
remain smaller than the scalar-hair-free SAdS solution independently of the choice
for the coupling function $f(\phi)$ - this is in complete agreement with the profile
found in the asymptotically-flat case \cite{ABK}. This behaviour persists for even
larger values of the negative cosmological constant for all classes of solutions
apart from the one emerging for the logarithmic function whose horizon area
is significantly increased in the small-mass regime, as may be seen from the
right plot of Fig. \ref{Area}. These plots verify also the
termination of all branches of solutions at the point of a minimum horizon, or
minimum mass, that all our GB solutions exhibit as a consequence of the
inequality (\ref{C-def}). We also observe that, as hinted by the small-$\Lambda$
approximation given in Eq. (\ref{C-newdef}), an increase in the value of the 
negative cosmological constant pushes upwards the lowest allowed value of the
horizon radius of our solutions.

%%%%%%%%%%%%%%%%%%%%%
%%%%%%%%%%%%%%%%%

%%%%%%%%%%%%%%%%%%%%%
\begin{figure}[t!]
\begin{center}
\minipage{0.51\textwidth} \hspace*{-0.4cm}
  \includegraphics[width=\linewidth]{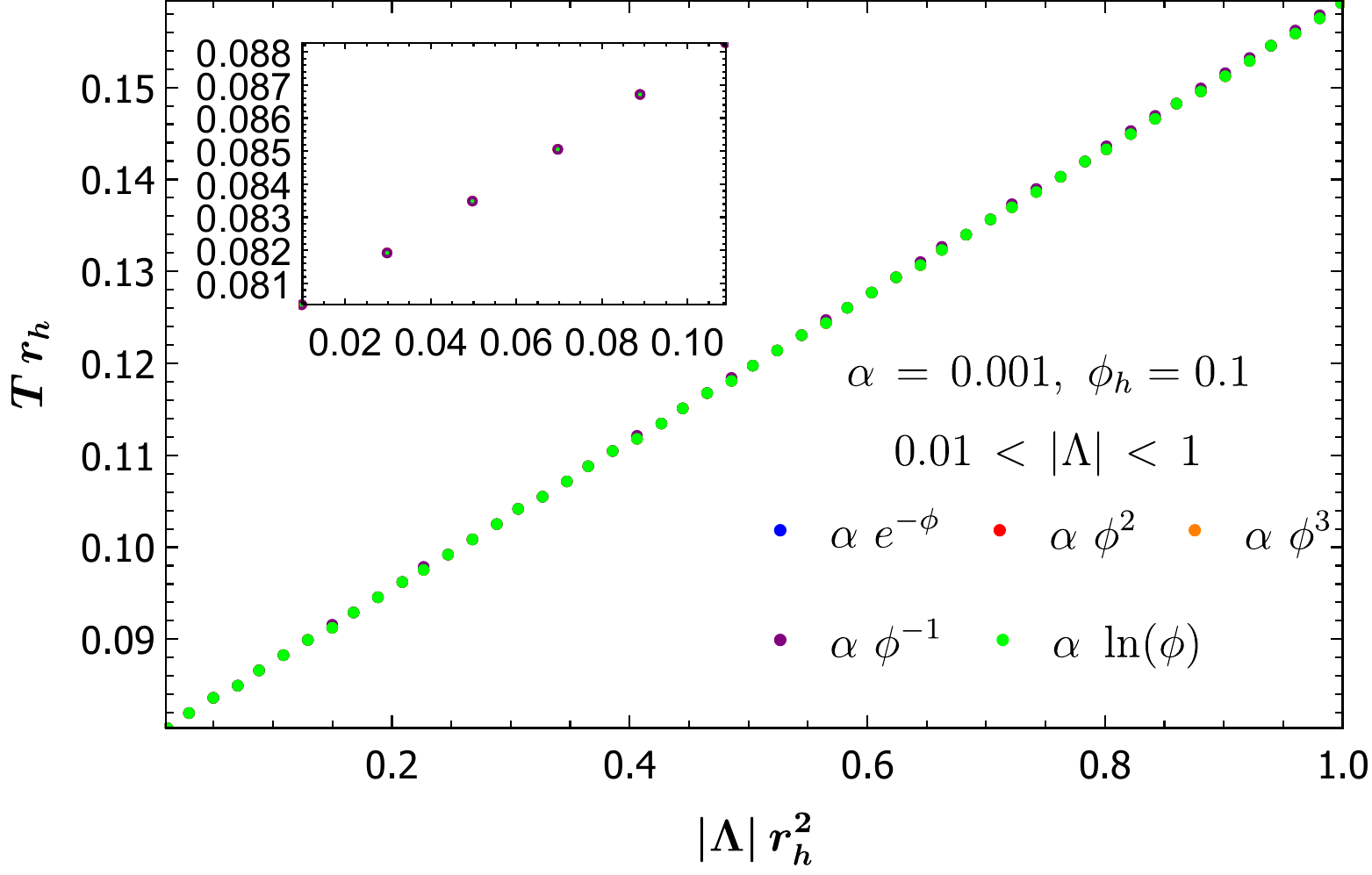}
\endminipage\hfill \hspace*{-1.5cm}
\minipage{0.51\textwidth}
  \includegraphics[width=\linewidth]{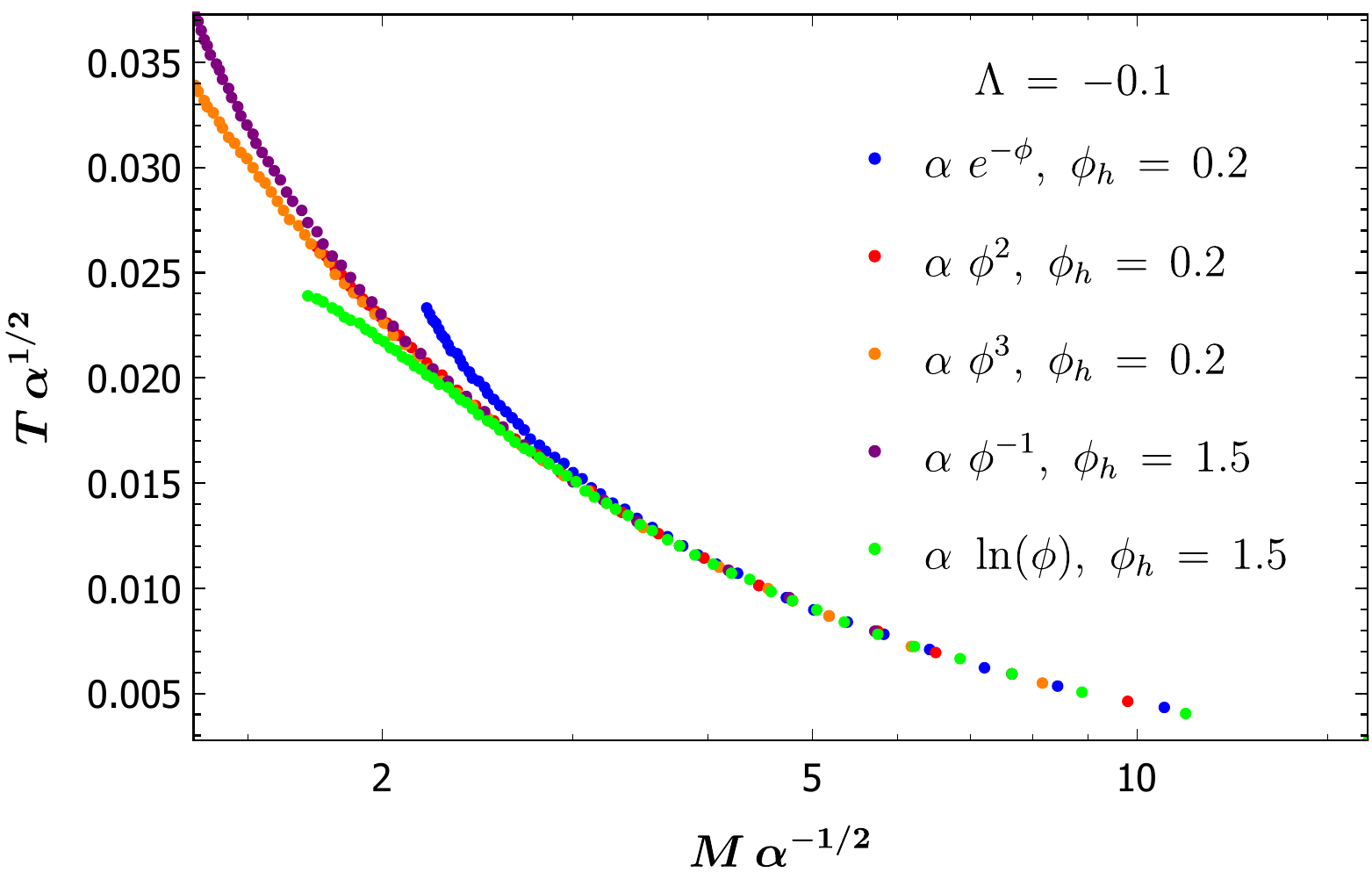}
\endminipage\hfill
    \caption{The temperature  $T$ of the black hole as a function of the cosmological
constant $\Lambda$ (left plot) and the mass $M$ of the black hole (right plot), for
various forms of $f(\phi)$.}
   \label{Temp-plot}
  \end{center}
\end{figure}
%%%%%%%%%%%%%%%%%

We now move to the thermodynamical quantities of our black-hole solutions. We start
with their temperature $T$ given by Eq. (\ref{Temp-def}) in terms of the near-horizon
coefficients $(a_1,b_1)$.  In the left plot of Fig. \ref{Temp-plot}, we display its
dependence in terms of the cosmological constant $\Lambda$, for several forms
of the coupling function. We observe that $T$ increases, too, with $|\Lambda|$;
we thus conclude that the more negatively-curved the spacetime is, the hotter
the black hole, that is formed, is. Note that the form of the coupling function
plays almost no role in this relation with the latter thus acquiring a universal 
character for all GB black-hole solutions. The dependence of the temperature
of the black hole on its mass, as displayed in the right plot of
Fig. \ref{Temp-plot}, exhibits a decreasing profile, with the obtained
solution being colder the larger its mass is. For small black-hole solutions,
the exact dependence of $T$ on $M$ depends on the particular form of the
coupling function but for solutions with a large mass its role becomes unimportant
as a common `Schwarzschild-AdS regime' is again approached. 

%%%%%%%%%%%%%%%%%%%%%
\begin{figure}[b!]
\begin{center}
\minipage{0.49\textwidth} \hspace*{-0.2cm}
  \includegraphics[width=\linewidth]{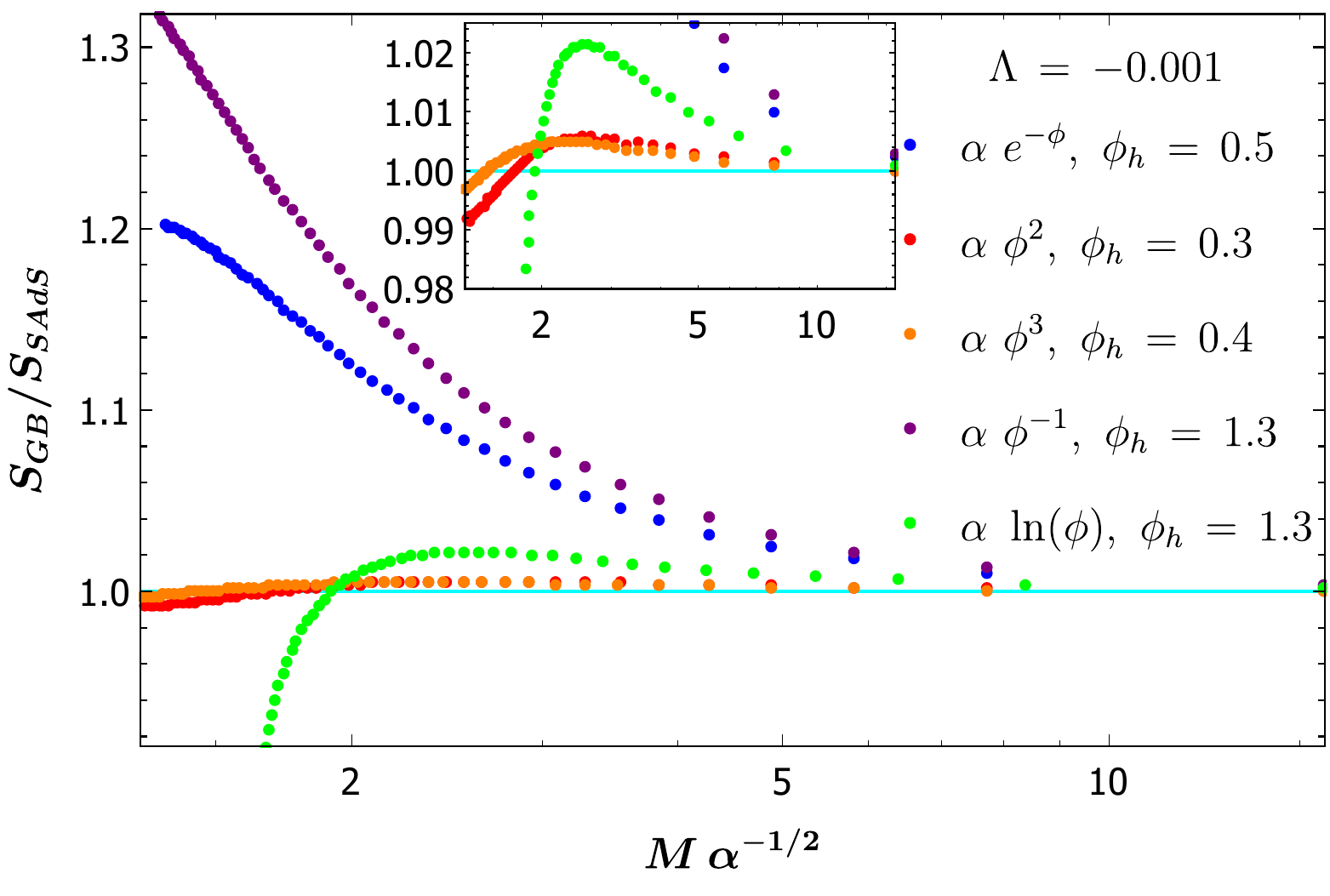}
\endminipage\hfill \hspace*{-1.8cm}
\minipage{0.49\textwidth}
  \includegraphics[width=\linewidth]{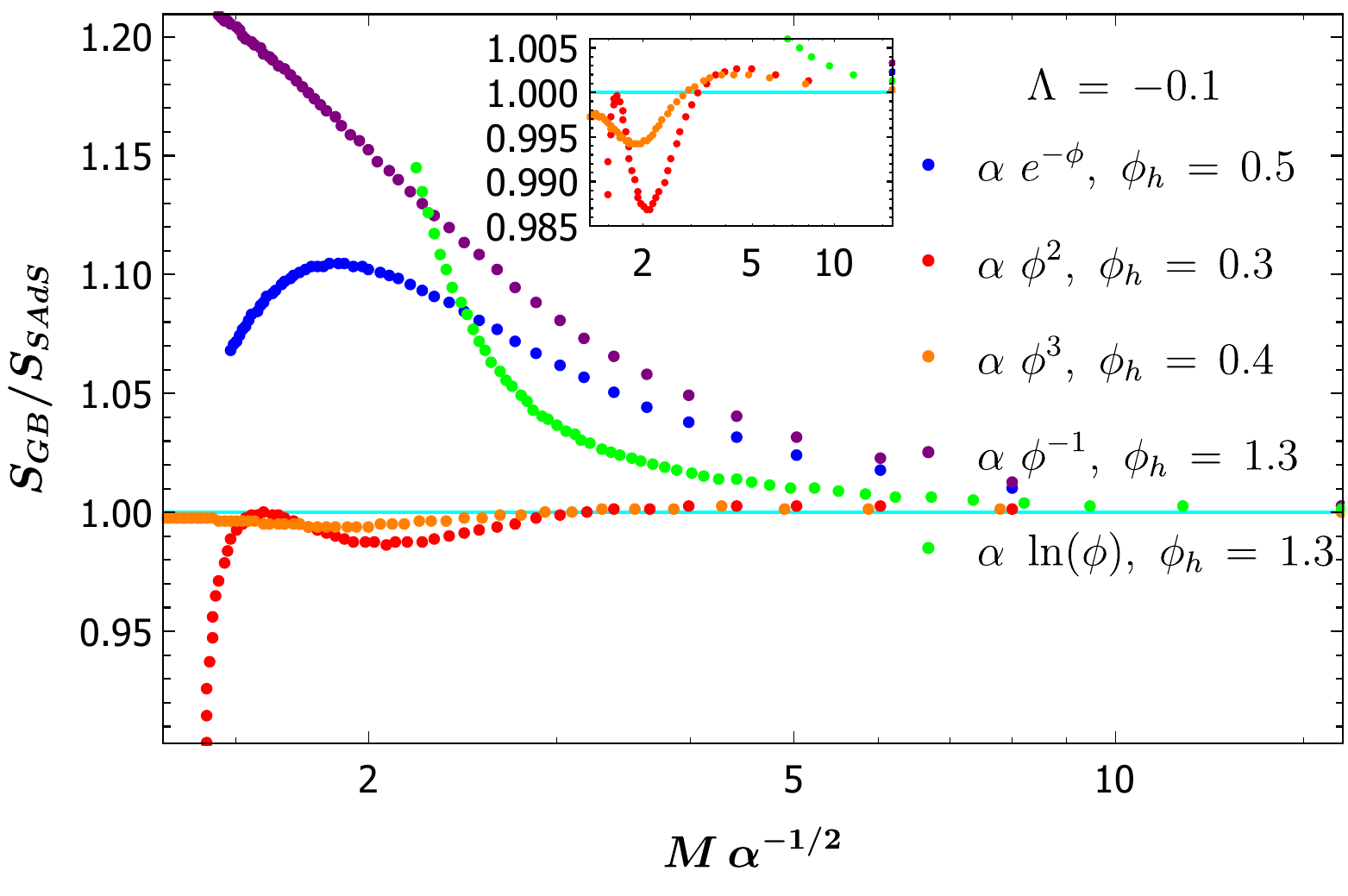}
\endminipage\hfill
    \caption{The entropy ratio $S_{GB}/S_{SAdS}$ of our solutions as a function of the
mass $M$ of the black hole, for various forms of $f(\phi)$, and for $\Lambda=-0.001$
(left plot) and $\Lambda=-0.1$ (right plot).}
   \label{S}
  \end{center}
\end{figure}
%%%%%%%%%%%%%%%%%

Let us finally study the entropy of the derived black-hole solutions. In Fig. \ref{S}, we
display the ratio of the entropy of our GB solutions over the entropy of the corresponding
Schwarzschild-Anti-de Sitter solution with the same mass, for the same indicative values
of the negative cosmological constant as for the horizon area. i.e. for
$\Lambda=-0.001$ (left plot) and $\Lambda=-0.1$ (right plot).
We observe that the profile of this quantity depends strongly on the choice of the coupling
function $f(\phi)$, for solutions with small masses, whereas in the limit of large mass,
where our solutions reduce to the SAdS ones, this ratio approaches unity as expected.
For small values of $\Lambda$, the left plot of Fig. \ref{S} depicts a behaviour similar
to the one found in the asymptotically-flat case \cite{ABK}: solutions emerging for the
linear and the quadratic coupling functions exhibit smaller entropy compared to the
SAdS one, while solutions for the exponential, logarithmic and inverse-linear coupling
functions lead to GB black holes with a larger entropy over the whole mass range or
for particular mass regimes. As we increase the value of the cosmological constant
(see right plot of Fig. \ref{S}), the entropy ratio is suppressed for all families of GB
black holes apart from the one emerging for the logarithmic coupling function, which
exhibits a substantial increase in this quantity over the whole mass regime. Together
with the solutions for the exponential and inverse-linear coupling functions, they have
an entropy ratio larger than unity while this ratio is now significantly lower than
unity for all the other polynomial coupling functions.
Although the question of the stability of the derived solutions
is an important one and must be independently studied for each family of solutions
found, the entropy profiles presented above may provide some hints regarding the
thermodynamical stability of our solutions compared to the Schwarzschild-Anti-de Sitter ones. 

%%%%%%%%%%%%%%%%%%%%%%%%%%%%%%%%%%%%%%%%%%%%%%%%%%%%%%%%%%%%%%%%%%%%%%

\subsection{de Sitter Gauss-Bonnet Black Holes}

We now address the case of a positive cosmological constant, $\Lambda>0$. We start our
integration process at a distance close to the black-hole horizon, using the asymptotic
solutions (\ref{A-rh})-(\ref{phi-rh}) and choosing $\phi_h$ to satisfy again the regularity
constraint (\ref{solf}). The coupling function $f(\phi)$ is assumed to take on a variety
of forms -- namely exponential, even and odd polynomial, inverse even and odd
polynomial, and logarithmic forms -- as in the case of the negative cosmological constant.
The numerical integration then proceeds outwards to meet the corresponding asymptotic
solution (\ref{A-rc})-(\ref{phi-rc}) near the cosmological horizon. 

Unfortunately, and despite our persistent efforts, no complete black-hole solution
interpolating between the asymptotic solutions (\ref{A-rh})-(\ref{phi-rh}) and 
(\ref{A-rc})-(\ref{phi-rc}) was found. The same negative result concerning the
existence of a black hole solution with an asymptotically de Sitter behaviour was
obtained in \cite{Hartmann}, where the case of a linear coupling function between
the GB term and the scalar field was
considered. It is, however, worth noting that the two asymptotic solutions near
the black-hole and cosmological horizons do independently emerge -- it is the
effort to match them in a smooth way via an intermediate solution that fails.

%%%%%%%%%%%%%%%%%%%%%
\begin{figure}[t!]
\begin{center}
\minipage{0.495\textwidth} \hspace*{-0.4cm}
  \includegraphics[width=\linewidth]{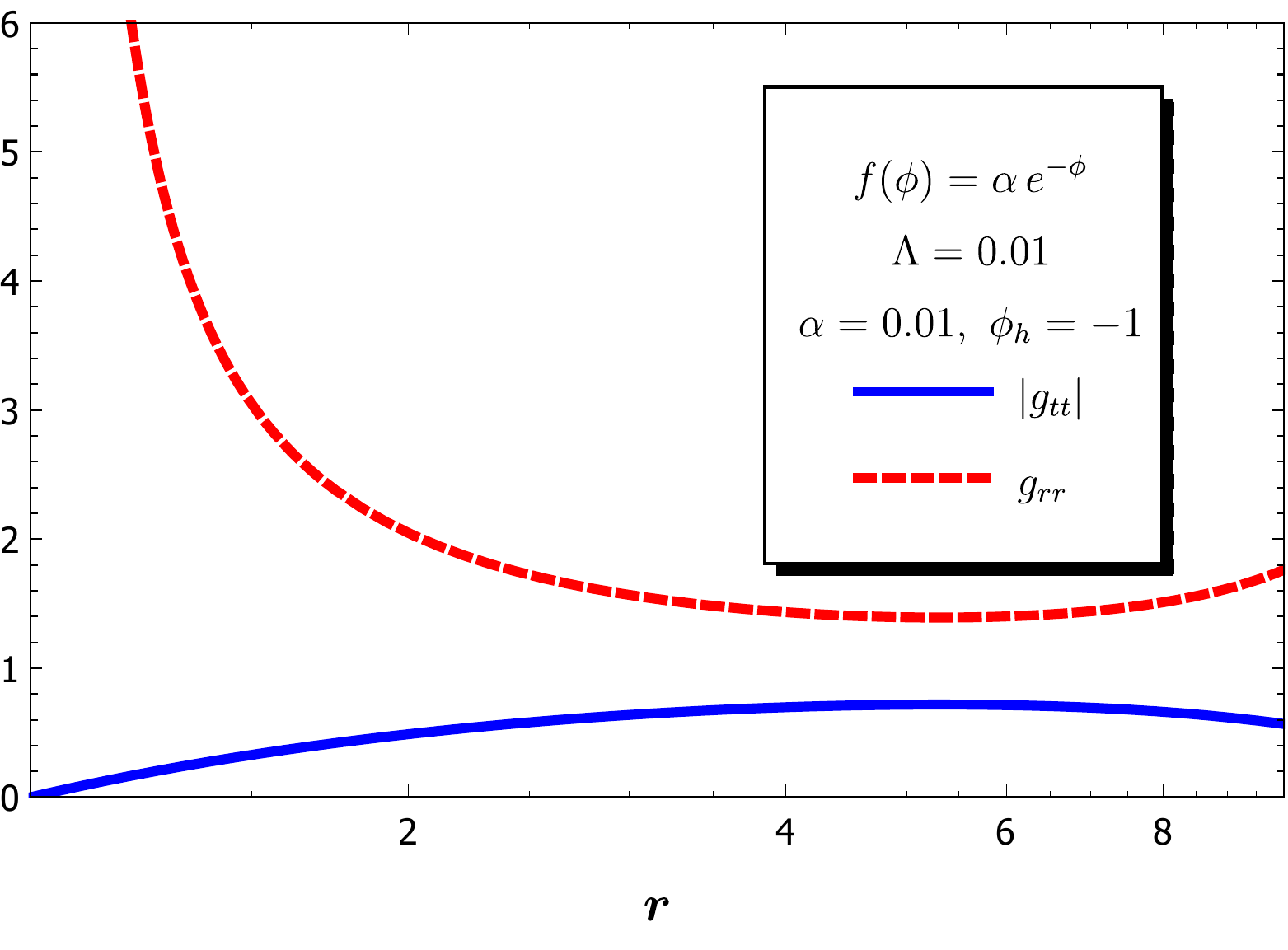}
\endminipage\hfill \hspace*{-1.0cm}
\minipage{0.51\textwidth}
  \includegraphics[width=\linewidth]{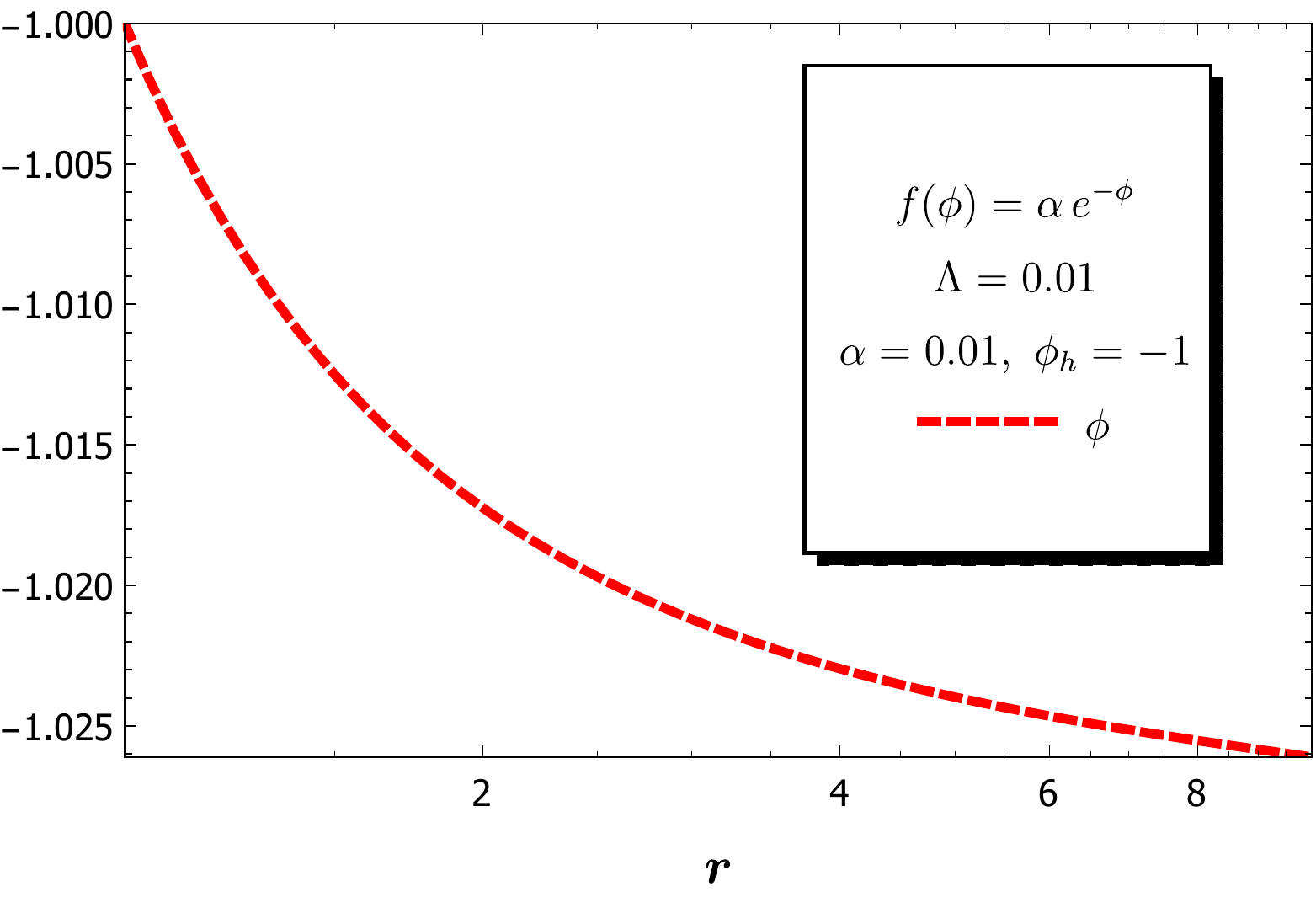}
\endminipage\hfill
    \caption{The metric functions  $|g_{tt}|$ and $g_{rr}$ of the spacetime (left plot)
and the scalar field $\phi$ (right plot) in terms of the radial coordinate $r$, for
a positive cosmological constant and coupling function $f(\phi)=\alpha e^{-\phi}$.}
   \label{metric-phi-dS}
  \end{center}
\end{figure}
%%%%%%%%%%%%%%%%%

To demonstrate this, in Fig. \ref{metric-phi-dS} we display the result of our numerical
integration for the indicative case of $\alpha=0.01$, $\phi_h=-1$ and $\Lambda=0.01$.
The coupling function has been chosen to be $f(\phi)=\alpha e^{-\phi}$, however, the
same qualitative behaviour was found for every choice of $f(\phi)$ we have considered. 
From the metric functions and the scalar-field profiles displayed in the two plots, we
clearly see that an asymptotic solution describing a regular black-hole horizon is indeed
formed. In this, the metric component $|g_{tt}|$ vanishes while the $g_{rr}$ one
diverges, as expected. The scalar field near the black-hole horizon assumes a finite,
constant value while it decreases away from the horizon, in perfect agreement with
the scalar-field profile found in the case of a negative cosmological constant. The
integration proceeds uninhibited but stops abruptly close to the regime where the
cosmological horizon should form. In fact, from the left plot of Fig. \ref{metric-phi-dS},
we may clearly see the expected behaviour of the metric components near the
cosmological horizon (i.e. the vanishing of $|g_{tt}|$ and divergence of $g_{rr}$)
just to emerge. 

The emergence of asymptotic solutions and the failure to smoothly match them
strongly reminds us of the analysis involved in the no-hair theorems \cite{NH-scalar,
Bekenstein}, where a similar situation holds. It is, however, difficult to generalise
that analysis, or equivalently the argument for their evasion as developed in
\cite{ABK}, in the present case of a non-vanishing cosmological constant \footnote{A
theoretical analysis is currently under way but has, so far, not given any conclusive
results.}. One could, nevertheless, gain some understanding of the situation by
examining the form of the near-horizon value of the $T^r_{\,\,r}$ component of
the energy-momentum tensor given in Eq. (\ref{Trr_rh}) 
-- the profile of this component is of paramount importance for the evasion of the
novel no-hair theorem \cite{Bekenstein} and the emergence of novel solutions. For
the evasion to be realised, this component must be positive and decreasing close
to the black-hole horizon \cite{DBH, ABK}. From Eq. (\ref{Trr_rh}), it becomes clear
that the presence of a negative cosmological constant ($\Lambda <0$) in the theory
always gives a positive contribution to $T^r_{\,\,r}$, and enhances the probability of
obtaining regular black holes. This justifies the easiness in which novel black-hole
solutions with an asymptotically Anti-de Sitter behaviour have emerged in the context
of our analysis. On the other hand, the contribution of a positive cosmological constant
($\Lambda >0$) to  $T^r_{\,\,r}$ is always negative, and this makes the evasion of
the no-hair theorem less likely. It would be indeed interesting to re-address the 
arguments presented in \cite{ABK} as well as the ones employed in the versions
of the no-hair theorems for non-asymptotically-flat black holes 
\cite{Narita, Winstanley-Nohair, Bhatta2007, Bardoux, Kazanas} to cover also the
case where the GB term and the cosmological constant appear simultaneously in the
theory. 

Nevertheless, even if the evasion of the no-hair theorems may be realised for
$\Lambda>0$ in the presence of the GB term 
-- for small values of $\Lambda$ this seems quite likely -- this merely opens the way
to look for novel solutions, it does not guarantee their existence. The emergence of a
complete solution interpolating between the two horizons still demands the smooth matching
of the two asymptotic solutions. It is quite likely that the system does not have enough
freedom to simultaneously satisfy the requirements for the existence of a regular solution,
namely Eqs. (\ref{solf})-(\ref{C-def}) and  (\ref{solfc})-(\ref{tildeC-def}) - this was also noted
in \cite{Hartmann}. Or, that a very careful selection of parameters may be necessary for
such a solution to emerge. In any case, further investigation is necessary, and we hope to
return to this topic soon. 

%%%%%%%%%%%%%%%%%%%%%%%%%%%%%%%%%%%%%%%%%%%%%%%%%%%%%%%%%%%%%%%%%%%%%

\section{Conclusions} 

In this work, we have extended our previous analyses \cite{ABK}, on the emergence of
novel, regular black-hole solutions in the context of the Einstein-scalar-GB theory, to include
the presence of a positive or negative cosmological constant. Since the uniform distribution
of energy associated with the cosmological constant permeates the whole spacetime, we
expected $\Lambda$ to have an effect on both the near-horizon and far-field asymptotic
solutions. Indeed, our analytical calculations in the small-$r$ regime revealed that the 
cosmological constant modifies the constraint that determines the value of $\phi'_h$ for
which a regular, black-hole horizon forms. In addition, it was demonstrated that
such a horizon is indeed formed, for either positive or negative $\Lambda$ and for all
choices of the coupling function $f(\phi)$. 

In contrast, the behaviour of the solution in the far-field regime depended strongly on
the sign of the cosmological constant. For $\Lambda>0$, a second horizon, the cosmological
one, was expected to form at a distance $r_c>r_h$, whereas for $\Lambda<0$, an Anti-de Sitter
type of solution was sought for at asymptotic infinity. Both types of solutions were analytically
shown to be admitted by the set of our field equations at the limit of large distances, thus
opening the way for the construction of complete black-hole solutions with an 
(Anti)-de Sitter asymptotic behaviour. 

The complexity of the field equations prevented us from constructing such a solution
analytically, therefore we turned to numerical analysis. Using our near-horizon analytic
solution as a starting point, we integrated the set of field equations from the black-hole
horizon and outwards. For a negative cosmological constant ($\Lambda<0$), we 
demonstrated that regular black-hole solutions with an Anti-de Sitter-type asymptotic
behaviour arise with the same easiness that their asymptotically-flat counterparts emerge. 
We have produced solutions for an exponential, polynomial (even or odd), inverse
polynomial (even or odd) and logarithmic coupling function between the scalar field and
the GB term. In each and every case, once $f(\phi)$ was chosen, selecting the input
parameter $\phi_h'$ to satisfy the regularity constraint (\ref{solf}) and the second
input parameter $\phi_h$ to satisfy the inequality (\ref{C-def}) a regular black hole
solution always emerged. The metric components exhibited the expected behaviour near
the black-hole and asymptotic infinity with the scalar invariant GB term being everywhere
regular. All solutions possessed non-trivial scalar hair, with the scalar field having a
non-trivial profile both close to and far away from the black-hole horizon. For small
negative values of $\Lambda$, we recovered the power-law fall-off of the scalar field
at infinity, found in the asymptotically-flat case \cite{ABK} whereas for large
negative values of $\Lambda$ the profile of $\phi$ was dominated by a logarithmic
dependence on the radial coordinate. This behaviour was analytically shown to emerge
both in the linear coupling-function case and in the perturbative limit, in terms of the
coupling parameter $\alpha$, but it was numerically found to accurately describe all
of our solutions at large distances.  

The absence of a $(1/r)$-term in the expression of the scalar field at large
distances excludes the presence of a scalar charge, even a secondary one. The 
coefficient $d_1$ in front of the logarithmic term in the expression of $\phi$ can
give us information on how much the large-distance behaviour of the scalar
field deviates from the power-law one valid in the asymptotically-flat case. We have
found that this deviation is stronger for GB black holes with a small mass 
whereas the more massive ones have a $d_1$ coefficient that tends to zero.
The temperature of the black holes was found to increase with the cosmological
constant independently of the form of the coupling function. The latter plays a
more important role in the relation of $T$ with the black-hole mass: while 
the temperature decreases with $M$ for all classes of solutions found, the
lighter ones exhibit a stronger dependence on $f(\phi)$. The same dependence
on the form of the coupling function is observed in the entropy and horizon area
of our solutions. For small masses, the entropy of each class of solutions has 
a different behaviour, with the ones for the exponential, inverse-linear polynomial and
logarithmic coupling functions exhibiting a ratio $S_{GB}/S_{SAdS}$ (over the
entropy of the Schwarzschild-Anti-de Sitter black hole with the same mass)
larger than unity for the entire mass range, for large values of $\Lambda$.
This feature hints towards the enhanced thermodynamical stability of our
solutions compared to their General Relativity (GR) analogues.
In the limit of large mass, the entropy of all classes of our solutions tend to the one of 
the Schwarzschild-Anti-de Sitter black hole with the same mass. The same holds
for the horizon area: while for small masses, each class has its own pattern
with $M$, will all solutions being smaller in size than the corresponding SAdS one
apart from the logarithmic case, for large masses all black-hole solutions match
the horizon area of the SAdS solution. 

Based on the above, we conclude that our GB black-hole solutions with a negative
cosmological constant smoothly merge with the SAdS ones, in the large mass limit. 
As in the asymptotically-flat case, it is the small-mass range that provides
the characteristic features for the GB solutions. These solutions have a modified
dependence of both their temperature and horizon area on their mass compared to
the SAdS solution. Another characteristic is also the minimum horizon, or minimum
mass, that all our GB solutions possess due to the inequality (\ref{C-def}). 

Turning to GB solutions with a positive cosmological constant, our quest has failed
to find any such solutions. Although the presence of a positive $\Lambda$ does not
obstruct the formation of a regular black-hole or cosmological horizon,
our numerical integration did not manage to produce a complete solution that would
interpolate between the two asymptotic regimes. This result holds independently of the
choice of the GB coupling function $f(\phi)$ or the value of $\Lambda$. 

In conclusion, we have demonstrated that the general classes of theories that contain
the GB term and lead to novel black-hole solutions, continue to do so even in the
presence of a negative cosmological constant in the theory. In contrast, the presence
of a positive cosmological constant presents a severe obstacle for the formation of these
solutions. A further investigation is clearly necessary in both cases: the relevance of the
GB solutions with an Anti-de Sitter-type asymptotic solution in the context of the AdS-CFT
correspondence should be inquired, and the deeper reason for the absence of solutions
with a positive cosmological constant should be investigated further. We hope to return
soon with results on both issues.

%%%%%%%%%%%%%%%%%%%%%%%%%%%%%%%%%%%%%%%

{\bf Acknowledgments}
G.A. would like to thank Onassis Foundation for the financial
support provided through its scholarship program. This research is
co-financed by Greece and the European Union (European Social Fund- ESF)
through the Operational Programme ``Human Resources Development, Education and
Lifelong Learning'' in the context of the project ``Strengthening Human
Resources Research Potential via Doctorate Research'' (MIS-5000432),
implemented by the State Scholarships Foundation (IKY).

\appendix

\numberwithin{equation}{section}

\section{Set of Differential Equations}
\label{Equations}

Here, we display the explicit expressions of the coefficients $P$, $S$ and $Q$ that
appear in the system of differential equations (\ref{A-sys})-(\ref{phi}) whose solution determines
the metric function $A$ and the scalar field $\phi$. Note, that in these
expressions we have eliminated, via Eq. (\ref{B'}), $B'$, that involves
$A''$ and $\phi''$, but retained $e^B$ for notational simplicity. They are:
%%%%%%%%%%%%%%%
\begin{align}
P&=-128 e^{4 B} \Lambda ^2 r^3 \dot f \left(r A'+2 e^B-2\right)+16
   A'^3 \dot f \Bigl[     
   -2 e^B \left(-14 e^B+3 e^{2 B}+19\right)
   r \dot f \phi ' \nn \\
   &+8\left(-8 e^B+3 e^{2 B}+9\right) \dot f^2
   \phi '^2-e^{2 B} \left(3 e^B-5\right) r^2  \Bigr] 
   +4 e^B  A'^2 \bigg\{
   e^B r \dot f \Bigl[\left(5 e^B-19\right) r^2
    \phi'^2 \nn\\
    &+12 \left(e^B-1\right)^2\Bigr]
  -4 \dot f^2 \phi' \Big[\left(9 e^B-17\right) r^2
   \phi'^2+8 \left(e^B-1\right)^2\Big] +e^{2 B} r^4
   \phi '\bigg\}\nn\\  
  &+ 4 e^{2B} 2 \,\L \biggl\{-e^{2B} r^3 (-2+rA') \f' -16 A' \dot f^2 \f' \left[ 6(3-4e^B +e^{2B}) +(-5 +e^B)r A' \right] \nn\\
  &+ 4 e^B \dot f \Bigl[ -3 r^2 A'^2(1+e^B) +4 \Bigl(4(-1+e^B)^2-r^2\f'^2\Bigr)+2rA'(3-3e^B+r^2\f'^2)\Bigr]\biggr\} \nn\\
  &-2e^{2B} r \f' \biggl\{-8\dot f \f' \Big[4 e^B (-1+e^B) + r^2 \f'^2 (-2+e^B)\Big]
 -4re^B(-1+e^B) \nn\\
  &-r\f'^2\Bigl[ r^2e^B-16\ddot f(-1+e^B) \Bigr]\biggr\}
  - A' e^B \biggl\{ 32 r \dot f^2 \f^2 \f' (9-4e^B+3e^{2B}) \nn\\
  &-r^3\f' e^B\Big[4e^B(1+e^B)-\f'^2\Bigl(r^2 e^B+16\ddot f (1+e^B)\Big) \Big]\nn\\
  &+8 e^B \dot f \Big[ 4(-1+e^B)^2  +r^2 \f'^2(-7+3e^B)
  -2\f'^4(r^4 +8r^2\ddot f)\Big]\bigg\},
  \end{align}
%%%%%%%%%%%%%%%%%%%%%%%
\begin{align}
S&=2304 A'  \dot f^3 \phi'^2 + 8 e^B \Big[  -128 r A'  \dot f^2 \phi '-448 A' \dot f^3
   \phi'^2+32 r^2 \dot f^2 \phi'^3-80 \dot f^2 \phi' \Big] \nn\\
   &+8e^{2B} \Big[ 16 r^2 A' f'+160 r A' \dot f^2 \phi '+160 A'
   \dot f^3 \phi '^2-12 r^3 \dot f \phi'^2 -16 r^2 \dot f^2 \phi '^3\nn\\
   &-64 \Lambda  r^2 \dot f^2 \phi '+16 r \dot f+160
   \dot f^2 \phi ' \Big] + 8 e^{3B} \Big[ -16 r^2 A' \dot f
   -32 r A' \dot f^2 \phi '+4 r^3 \dot f \phi '^2 \nn\\
   &+16 \Lambda  r^3 \dot f+64 \Lambda  r^2 \dot f^2 \phi '-32 r \dot f
   -80 \dot f^2 \phi
   '+r^4 \phi ' \Big] +8 e^{4B} \Big[  16 r \dot f-16 \Lambda  r^3 \dot f \Big],
\end{align}
%%%%%%%%%%%%%%%
and
%%%%%%%%%%%%%%%%%
\begin{align}
Q&=2304 A'  \dot f^2 \ddot f  \phi'^4-1152
   A'^2 \dot f^3 \phi'^3 + e^B \Big[-144 r^2 A' \dot f^2 \phi'^4+672 r
   A'^2 \dot f^2 \phi '^2\nn\\
   &+768 A'^2 \dot f^3 \phi '^3-384 A'
   \dot f^2 \phi '^2-1024 r A' \dot f \ddot f
   \phi '^3-3584 A' \dot f^2 \ddot f \phi
   '^4\nn\\
   &+480 r \dot f^2 \phi '^4+64 r^2
   \dot f \ddot f \phi '^5-640 \dot f \ddot f \phi '^3 \Big] 
   + e^{2B} \Big[ 128 r^2 A' \ddot f \phi '^2+52 r^3 A' \dot f \phi
   '^3\nn\\
   &+80 r^2 A' \dot f^2 \phi '^4-128
   r^2 A'^2 \dot f \phi '-576 \Lambda  r^2 A'
   \dot f^2 \phi '^2-320 r A'^2
   \dot f^2 \phi '^2\nn\\
   &+176 r A' \dot f \phi '-128
   A'^2 \dot f^3 \phi '^3+640 A'
   \dot f^2 \phi '^2+1280 r A' \dot f \ddot f
   \phi '^3 +1280 A' \dot f^2 \ddot f \phi'^4\nn\\
   &-16 r^3 \ddot f \phi '^4+128 r \ddot f
   \phi '^2-4 r^4 \dot f \phi '^5-152 r^2
   \dot f \phi '^3 -256 r \dot f^2 \phi
   '^4+384 \Lambda  r \dot f^2 \phi'^2\nn\\
   &+160 \dot f \phi '-64 r^2 \dot f \ddot f \phi'^5
   -512 \Lambda  r^2 \dot f \ddot f \phi '^3+1280
   \dot f \ddot f \phi '^3\Big] + e^{3B} \Big[ -128 r^2 A' \ddot f \phi '^2\nn\\
   &+208 \Lambda  r^3 A' \dot f \phi '+32 r^2
    A'^2 \dot f \phi '+320 \Lambda  r^2 A'
   \dot f^2 \phi '^2+32 r A'^2
   \dot f'^2 \phi '^2-224 r A' \dot f \phi '\nn\\
   &-256
   A' \dot f^2 \phi '^2-256 r A' \dot f \ddot f
   \phi '^3-6 r^4 A' \phi '^2+8 r^3
   A'^2-24 r^2 A'+16 r^3 \ddot f \phi '^4\nn\\
   &+128 \Lambda  r^3 \ddot f \phi '^2-256 r \ddot f \phi
   '^2+16 \Lambda  r^4 \dot f \phi '^3+24 r^2 \dot f \phi
   '^3 -12 r^3 A' \dot f \phi'^3 +32 r \dot f^2 \phi '^4\nn\\
   &+224 \Lambda  r^2 \dot f \phi '-512 \Lambda  r \dot f^2
   \phi '^2-320 \dot f \phi '
   +512 \Lambda  r^2 \dot f \ddot f
   \phi '^3-640 \dot f \ddot f \phi '^3+r^5
   \phi '^4\nn\\
   &+12 r^3 \phi '^2-32 r \Big] +e^{4B} \Big[ -48 \Lambda  r^3 A' \dot f \phi '
   +48 r A' \dot f \phi '-24 \Lambda  r^4
   A'+24 r^2 A'\nn\\
   &-128 \Lambda  r^3 \ddot f \phi '^2+128 r
   \ddot f \phi '^2+128 \Lambda ^2 r^4 \dot f \phi'
   -224 \Lambda  r^2 \dot f
   \phi '+128 \Lambda  r \dot f^2 \phi
   '^2+160 \dot f \phi '\nn\\
   &-4 \Lambda  r^5 \phi'^2+4 r^3 \phi
   '^2-64 \Lambda  r^3+64 r \Big] +e^{5B} \Big[  -32 \Lambda ^2 r^5+
   64 \Lambda  r^3-32 r \Big].
\end{align}
%%%%%%%%%%%%%%%%

%%%%%%%%%%%%%%%%%%%%%%%%%%%%%%%%%%%%%%%%%%%%%%%%%%%%%%%%%%%

\section{Scalar Quantities}
\label{scalar}

By employing the metric components of the line-element (\ref{metric}), one may
compute the following scalar-invariant gravitational quantities:
%%%%%%%%%%%%%%%%%%%%%%
\bea
R&=&+\frac{e^{-B}}{2r^2}\left(4e^B-4-r^2A'^2+4rB'-4rA'+r^2A'B'-2r^2A''\right),\label{A1}\\\nonumber\\
R_{\mu\nu}R^{\mu\nu}&=&+\frac{e^{-2B}}{16 r^4}\left[8(2-2e^B+rA'-rB')^2+r^2(rA'^2-4B'-rA'B'+2rA'')^2\right.\nonumber\\
&&\left.+r^2(rA'^2+A'(4-rB')+2rA'')^2\right],\\\nonumber\\
R_{\mu\nu\rho\sigma}R^{\mu\nu\rho\sigma}&=&+\frac{e^{-2B}}{4r^4}\left[r^4A'^4-2r^4A'^3B'-4r^4A'B'A''+r^2A'^2(8+r^2B'^2+4r^2A'')\right.\nonumber\\
&&\left.+16(e^B-1)^2+8r^2B'^2+4r^4A''^2\right],\\\nonumber\\
R_{GB}^2&=&+\frac{2e^{-2B}}{r^2}\left[(e^B-3)A'B'-(e^B-1)A'^2-2(e^B-1)A''\right].\label{A4}
   \eea
%%%%%%%%%%%%%%%%%%%%%%%%%%%%%%%%%%

\section{Variation with respect to the Riemann tensor}
\label{variation}

Here, we derive the derivatives of the Lagrangian of the theory (\ref{action}) with respect to
the Riemann tensor. A simple way to do this is to take the derivatives ignoring the symmetries,
that the final expression should possess, and restore them afterwards. For example, if
$A_{abcd}$ is a 4-rank tensor and $A$ the corresponding scalar quantity, we may write:
%%%%%%%%%%%%%%%%
\begin{align}
\frac{\partial A}{\partial A_{abcd}}&=\frac{\partial}{\partial A_{abcd}}\left(g^{\mu\rho}g^{\nu\sigma}A_{\mu\nu\rho\sigma}\right)=
g^{\mu\rho}g^{\nu\sigma}\delta^a_\mu\delta^b_\nu\delta^c_\rho\delta^d_\sigma=g^{ac}g^{bd}\,.
\end{align}
Now, if $A_{abcd}=R_{abcd}$, it should satisfy the following relations:
%%%%%%%%%%%%%
\begin{equation}\label{101}
A_{abcd}=A_{cdab}=-A_{abdc}\; \text{and}\; A_{abcd}+A_{acdb}+A_{adbc}=0.
\end{equation}
%%%%%%%%%%
Restoring the symmetries, we arrive at:
%%%%%%%%%%%%
\begin{equation}
\frac{\partial R}{\partial R_{abcd}}=\frac{1}{2}\left(g^{ac}g^{bd}-g^{bc}g^{ad}\right).
\end{equation}
Alternatively, we could have explicitly written:
\begin{align}
\frac{\partial R}{\partial R_{abcd}}&=\frac{\partial}{\partial R_{abcd}}\left(g^{\mu\rho}g^{\nu\sigma}R_{\mu\nu\rho\sigma}\right)=\frac{1}{2}g^{\mu\rho}g^{\nu\sigma}\frac{\partial}{\partial R_{abcd}}\left(R_{\mu\nu\rho\sigma}-R_{\nu\mu\rho\sigma}\right)\nonumber\\[1mm]
&=\frac{1}{2}g^{\mu\rho}g^{\nu\sigma}\left(\delta^a_\mu\delta^b_\nu\delta^c_\rho\delta^d_\sigma-\delta^a_\nu\delta^b_\mu\delta^c_\rho\delta^d_\sigma\right)=\frac{1}{2}\left(g^{ac}g^{bd}-g^{bc}g^{ad}\right),
\end{align}
%%%%%%%%%%%%
which clearly furnishes the same result. 

We now proceed to the higher derivative terms. Let us start with the Kretchmann scalar for which we find
%%%%%%%%%%%%%%%
\begin{equation}
\frac{\partial R_{\mu\nu\rho\sigma} R^{\mu\nu\rho\sigma}}{\partial R_{abcd}}=2R^{\mu\nu\rho\sigma}\frac{\partial R_{\mu\nu\rho\sigma}}{\partial R_{abcd}}=2R^{abcd},
\end{equation}
The above result does not need any correction as it is already proportional to $R_{abcd}$,
and satisfies all the desired identities. We now move to the $R_{\mu\nu} R^{\mu\nu}$
term, and employ again the simple method used above. Then: 
\begin{equation}
\frac{\partial A_{\mu\nu}A^{\mu\nu}}{\partial A_{abcd}}=2A^{\mu\nu}\frac{\partial A_{\mu\nu}}{\partial A_{abcd}}=2A^{\mu\nu}g^{\kappa\lambda}\frac{\partial A_{\kappa\mu\lambda\nu}}{\partial A_{abcd}}=g^{ac}A^{bd}-g^{bc}A^{ad},
\end{equation}
If $A_{abcd}=R_{abcd}$ and $A_{\mu\nu}=R_{\mu\nu}$, the above result will have all the right
properties if it is rewritten as
%%%%%%%%%%%%%%%%
\begin{equation}
\frac{\partial R_{\mu\nu}R^{\mu\nu}}{\partial R_{abcd}}=\frac{1}{2}\left(g^{ac}R^{bd}-g^{bc}R^{ad}-g^{ad}R^{bc}+g^{bd}R^{ac}\right),
\end{equation}
which is indeed the correct result. Finally, we easily derive that
\begin{equation}
\frac{\partial R^2}{\partial R_{abcd}}=R\left(g^{ac}g^{bd}-g^{bc}g^{ad}\right).
\end{equation}
%%%%%%%%%%%%%%%%%%%%%%%%%%%%%%%%%%%%%%%%%%%%%%%%%%%%%

In order to compute the integral appearing in Eq. (\ref{S2}), we use the near-horizon
solution (\ref{A-rh})-(\ref{phi-rh}) for the metric functions and scalar field. Then
recalling that, near the horizon, the relations $A''\approx -A'^2$ and $B'\approx -A'$
also hold, we find the results
%%%%%%%%%%%%%
\begin{align*}
R^{0101}\big|_{\mathcal{H}} = -\frac{1}{4} e^{-A-2 B} \left(-2 A''+A'
   B'-A'^2\right)\big|_{\mathcal{H}} \rightarrow 0\,, &\\[1mm]
-2\left(g^{00}R^{11}-g^{10}R^{01}-g^{01}R^{10}+g^{11}R^{00}\right)\big|_{\mathcal{H}}
\rightarrow \frac{4}{r_h}e^{-A-2B}A'\big|_\mathcal{H}\approx -\frac{4b_1^2}{a_1r_h}\,,& \\[1mm]
g^{00}g^{11}R\big|_{\mathcal{H}} \rightarrow \frac{e^{-A-2B}}{r^2}\left(4rA'-2e^B\right)\big|_{\mathcal{H}}\approx \frac{4b_1^2}{a_1r_h} -\frac{2b_1}{a_1 r_h^2}\,,& \\[2mm]
g_{00}g_{11}\big|_{\mathcal{H}}=e^{A+B}\big|_{\mathcal{H}}\rightarrow a_1/b_1\,.&
\end{align*}
%%%%%%%%%%%%%%%
Substituting these into Eq. (\ref{S2}), we readily obtain the result (\ref{S2-final}).
%%%%%%%%%%%%%%%%%%%%%%%%%%%%%%%%%%%%%%%%%%%%%%%%%%%%%%


\begin{thebibliography}{9}

\bibitem{Stelle} K.~S.~Stelle,
  %``Renormalization of Higher Derivative Quantum Gravity,''
  Phys.\ Rev.\ D {\bf 16} (1977) 953.
  %%CITATION = doi:10.1103/PhysRevD.16.953;%%
  
\bibitem{General} T.~P.~Sotiriou,
  %``Gravity and Scalar Fields,''
  Lect.\ Notes Phys.\  {\bf 892} (2015) 3;\\
  %%CITATION = doi:10.1007/978-3-319-10070-8_1;%%
%%%%%%%%%%%%%%%%%
E.~Berti {\it et al.},
  %``Testing General Relativity with Present and Future Astrophysical Observations,''
  Class.\ Quant.\ Grav.\  {\bf 32} (2015) 243001.
 
\bibitem{NH-scalar} J.~D.~Bekenstein,
  %``Transcendence of the law of baryon-number conservation in black hole physics,''
  Phys.\ Rev.\ Lett.\  {\bf 28} (1972) 452; 
  %%CITATION = doi:10.1103/PhysRevLett.28.452;%%
C.~Teitelboim,
  %``Nonmeasurability of the lepton number of a black hole,''
  Lett.\ Nuovo Cim.\  {\bf 3S2} (1972) 397.
  %%CITATION = doi:10.1007/BF02826050;%%

\bibitem{YM} M.~S.~Volkov and D.~V.~Galtsov,
 %``NonAbelian Einstein Yang-Mills black holes,''
 JETP Lett.\  {\bf 50} (1989) 346;
  %%CITATION = JTPLA,50,346;%% 
P.~Bizon,
%``Colored black holes,''
  Phys.\ Rev.\ Lett.\  {\bf 64} (1990) 2844;
  %%CITATION = doi:10.1103/PhysRevLett.64.2844;%%
B.~R.~Greene, S.~D.~Mathur and C.~M.~O'Neill,
  %``Eluding the no hair conjecture: Black holes in spontaneously broken gauge theories,''
  Phys.\ Rev.\ D {\bf 47} (1993) 2242;
  %%CITATION = doi:10.1103/PhysRevD.47.2242;%%
K.~i.~Maeda, T.~Tachizawa, T.~Torii and T.~Maki,
%``Stability of nonAbelian black holes and catastrophe theory,''
Phys.\ Rev.\ Lett.\  {\bf 72} (1994) 450.
  %%CITATION = doi:10.1103/PhysRevLett.72.450;%%

\bibitem{Skyrmions} H.~Luckock and I.~Moss,
  %``Black Holes Have Skyrmion Hair,''
  Phys.\ Lett.\ B {\bf 176} (1986) 341;
  %%CITATION = doi:10.1016/0370-2693(86)90175-9;%%
S.~Droz, M.~Heusler and N.~Straumann,
  %``New black hole solutions with hair,''
  Phys.\ Lett.\ B {\bf 268} (1991) 371.
  %%CITATION = doi:10.1016/0370-2693(91)91592-J;%%

\bibitem{Conformal} J.~D.~Bekenstein,
  %``Exact solutions of Einstein conformal scalar equations,''
  Annals Phys.\  {\bf 82} (1974) 535; 
  %%CITATION = doi:10.1016/0003-4916(74)90124-9;%%
Annals Phys.\  {\bf 91} (1975) 75.
  %%CITATION = doi:10.1016/0003-4916(75)90279-1;%%

\bibitem{Bekenstein} J.~D.~Bekenstein,
  %``Novel ¡¡no-scalar-hair¢¢ theorem for black holes,''
  Phys.\ Rev.\ D {\bf 51} (1995) no.12,  R6608.
  %%CITATION = doi:10.1103/PhysRevD.51.R6608;%%

\bibitem{DBH} P.~Kanti, N.~E.~Mavromatos, J.~Rizos, K.~Tamvakis and E.~Winstanley,
  %``Dilatonic black holes in higher curvature string gravity,''
  Phys.\ Rev.\ D {\bf 54} (1996) 5049; 
  %%CITATION = doi:10.1103/PhysRevD.54.5049;%%
Phys.\ Rev.\ D {\bf 57} (1998) 6255.
  %%CITATION = doi:10.1103/PhysRevD.57.6255;%%
  
\bibitem{Gibbons}
  G.~W.~Gibbons and K.~i.~Maeda,
  %``Black Holes and Membranes in Higher Dimensional Theories with Dilaton Fields,''
  Nucl.\ Phys.\ B {\bf 298} (1988) 741.
  %%CITATION = doi:10.1016/0550-3213(88)90006-5;%%
  
\bibitem{Callan}
  C.~G.~Callan, Jr., R.~C.~Myers and M.~J.~Perry,
  %``Black Holes in String Theory,''
  Nucl.\ Phys.\ B {\bf 311} (1989) 673.
  %%CITATION = doi:10.1016/0550-3213(89)90172-7;%%
  
\bibitem{Campbell}
  B.~A.~Campbell, M.~J.~Duncan, N.~Kaloper and K.~A.~Olive,
  %``Axion hair for Kerr black holes,''
  Phys.\ Lett.\ B {\bf 251} (1990) 34;
  %%CITATION = doi:10.1016/0370-2693(90)90227-W;%%
 B.~A.~Campbell, N.~Kaloper and K.~A.~Olive,
  %``Axion hair for dyon black holes,''
  Phys.\ Lett.\ B {\bf 263} (1991) 364.
  %%CITATION = doi:10.1016/0370-2693(91)90474-5;%%

\bibitem{Mignemi}
S.~Mignemi and N.~R.~Stewart,
  %``Charged black holes in effective string theory,''
  Phys.\ Rev.\ D {\bf 47} (1993) 5259.
  %%CITATION = doi:10.1103/PhysRevD.47.5259;%%
  
\bibitem{Kanti1995}
  P.~Kanti and K.~Tamvakis,
  %``Classical moduli O (alpha-prime) hair,''
  Phys.\ Rev.\ D {\bf 52} (1995) 3506.
  %%CITATION = doi:10.1103/PhysRevD.52.3506;%%

\bibitem{Torii}
  T.~Torii, H.~Yajima and K.~i.~Maeda,
  %``Dilatonic black holes with Gauss-Bonnet term,''
  Phys.\ Rev.\ D {\bf 55} (1997) 739.
  %%CITATION = doi:10.1103/PhysRevD.55.739;%%
  
\bibitem{KT}
  P.~Kanti and K.~Tamvakis,
  %``Colored black holes in higher curvature string gravity,''
  Phys.\ Lett.\ B {\bf 392} (1997) 30; 
  %%CITATION = doi:10.1016/S0370-2693(96)01521-3;%%
P.~Kanti and E.~Winstanley,
  %``Do stringy corrections stabilize colored black holes?,''
  Phys.\ Rev.\ D {\bf 61} (2000) 084032.
  %%CITATION = doi:10.1103/PhysRevD.61.084032;%%

\bibitem{Guo} Z.~K.~Guo, N.~Ohta and T.~Torii,
  %``Black Holes in the Dilatonic Einstein-Gauss-Bonnet Theory in Various Dimensions. I. Asymptotically Flat Black Holes,''
 Prog.\ Theor.\ Phys.\  {\bf 120} (2008) 581;
  %%CITATION = doi:10.1143/PTP.120.581;%%
 K.~i.~Maeda, N.~Ohta and Y.~Sasagawa,
  %``Black Hole Solutions in String Theory with Gauss-Bonnet Curvature Correction,''
  Phys.\ Rev.\ D {\bf 80} (2009) 104032;
  %%CITATION = doi:10.1103/PhysRevD.80.104032;%%
N.~Ohta and T.~Torii,
  %``Global Structure of Black Holes in String Theory with Gauss-Bonnet Correction in Various Dimensions,''
  Prog.\ Theor.\ Phys.\  {\bf 124} (2010) 207.
  %%CITATION = doi:10.1143/PTP.124.207;%%
  
 \bibitem{Kleihaus}
  B.~Kleihaus, J.~Kunz and E.~Radu,
  %``Rotating Black Holes in Dilatonic Einstein-Gauss-Bonnet Theory,''
  Phys.\ Rev.\ Lett.\  {\bf 106} (2011) 151104;
  %%CITATION = doi:10.1103/PhysRevLett.106.151104;%%
  B.~Kleihaus, J.~Kunz, S.~Mojica and E.~Radu,
  %``Spinning black holes in Einstein–Gauss-Bonnet–dilaton theory: Nonperturbative solutions,''
  Phys.\ Rev.\ D {\bf 93} (2016) no.4,  044047.
  %%CITATION = doi:10.1103/PhysRevD.93.044047;%%\bibitem{Mignemi:1992nt}
  
   \bibitem{Pani}
  P.~Pani, C.~F.~B.~Macedo, L.~C.~B.~Crispino and V.~Cardoso,
  %``Slowly rotating black holes in alternative theories of gravity,''
  Phys.\ Rev.\ D {\bf 84} (2011) 087501;
  %%CITATION = doi:10.1103/PhysRevD.84.087501;%%
  %%%%%%
  P.~Pani, E.~Berti, V.~Cardoso and J.~Read,
  %``Compact stars in alternative theories of gravity. Einstein-Dilaton-Gauss-Bonnet gravity,''
  Phys.\ Rev.\ D {\bf 84} (2011) 104035.
  %%CITATION = doi:10.1103/PhysRevD.84.104035;%%
  
\bibitem{Herdeiro}
  C.~A.~R.~Herdeiro and E.~Radu,
  %``Kerr black holes with scalar hair,''
  Phys.\ Rev.\ Lett.\  {\bf 112} (2014) 221101.
  %%CITATION = doi:10.1103/PhysRevLett.112.221101;%%
  
  \bibitem{Ayzenberg}
  D.~Ayzenberg and N.~Yunes,
  %``Slowly-Rotating Black Holes in Einstein-Dilaton-Gauss-Bonnet Gravity: Quadratic Order in Spin Solutions,''
  Phys.\ Rev.\ D {\bf 90} (2014) 044066
   Erratum: [Phys.\ Rev.\ D {\bf 91} (2015) no.6,  069905].
  %%CITATION = doi:10.1103/PhysRevD.91.069905, 10.1103/PhysRevD.90.044066;%%

\bibitem{Win-review} E.~Winstanley,
  %``Classical Yang-Mills black hole hair in anti-de Sitter space,''
  Lect.\ Notes Phys.\  {\bf 769} (2009) 49.
  %%CITATION = doi:10.1007/978-3-540-88460-6_2;%%

  \bibitem{Charmousis-rev}
  C.~Charmousis,
  %``Higher order gravity theories and their black hole solutions,''
  Lect.\ Notes Phys.\  {\bf 769} (2009) 299.
  %%CITATION = doi:10.1007/978-3-540-88460-6_8;%%
  
 \bibitem{Herdeiro-review}
  C.~A.~R.~Herdeiro and E.~Radu,
  %``Asymptotically flat black holes with scalar hair: a review,''
  Int.\ J.\ Mod.\ Phys.\ D {\bf 24} (2015) no.09,  1542014.
  %%CITATION = doi:10.1142/S0218271815420146;%%
  
   \bibitem{Blazquez}
  J.~L.~Blazquez-Salcedo {\it et al.},
  %``Black holes in Einstein-Gauß-Bonnet-dilaton theory,''
  IAU Symp.\  {\bf 324} (2017) 265.
  %%CITATION = doi:10.1017/S1743921316012965;%%

\bibitem{Metsaev} R.~R.~Metsaev and A.~A.~Tseytlin,
  %``Order alpha-prime (Two Loop) Equivalence of the String Equations of Motion and the Sigma Model %Weyl Invariance Conditions: Dependence on the Dilaton and the Antisymmetric Tensor,''
  Nucl.\ Phys.\ B {\bf 293} (1987) 385.
  %%CITATION = doi:10.1016/0550-3213(87)90077-0;%%
  
\bibitem{Horndeski} G.~W.~Horndeski,
  %``Second-order scalar-tensor field equations in a four-dimensional space,''
  Int.\ J.\ Theor.\ Phys.\  {\bf 10} (1974) 363.
  %%CITATION = doi:10.1007/BF01807638;%%

\bibitem{Galileon}
  A.~Nicolis, R.~Rattazzi and E.~Trincherini,
  %``The Galileon as a local modification of gravity,''
  Phys.\ Rev.\ D {\bf 79} (2009) 064036.
  %%CITATION = doi:10.1103/PhysRevD.79.064036;%%
  
\bibitem{SF} T.~P.~Sotiriou and V.~Faraoni,
  %``Black holes in scalar-tensor gravity,''
  Phys.\ Rev.\ Lett.\  {\bf 108} (2012) 081103.
  %%CITATION = doi:10.1103/PhysRevLett.108.081103;%%

\bibitem{HN}  L.~Hui and A.~Nicolis,
  %``No-Hair Theorem for the Galileon,''
  Phys.\ Rev.\ Lett.\  {\bf 110} (2013) 241104.
  %%CITATION = doi:10.1103/PhysRevLett.110.241104;%%
  
\bibitem{SZ}  T.~P.~Sotiriou and S.~Y.~Zhou,
  %``Black hole hair in generalized scalar-tensor gravity,''
  Phys.\ Rev.\ Lett.\  {\bf 112} (2014) 251102.
  %%CITATION = doi:10.1103/PhysRevLett.112.251102;%%
  
\bibitem{Babichev}
  E.~Babichev and C.~Charmousis,
  %``Dressing a black hole with a time-dependent Galileon,''
  JHEP {\bf 1408} (2014) 106.
  %%CITATION = doi:10.1007/JHEP08(2014)106;%%
   
\bibitem{Benkel}  T.~P.~Sotiriou and S.~Y.~Zhou,
  %``Black hole hair in generalized scalar-tensor gravity: An explicit example,''
  Phys.\ Rev.\ D {\bf 90} (2014) 124063;
  %%CITATION = doi:10.1103/PhysRevD.90.124063;%%
R.~Benkel, T.~P.~Sotiriou and H.~Witek,
  %``Black hole hair formation in shift-symmetric generalised scalar-tensor gravity,''
  Class.\ Quant.\ Grav.\  {\bf 34} (2017) no.6,  064001;
  %%CITATION = doi:10.1088/1361-6382/aa5ce7;%%
Phys.\ Rev.\ D {\bf 94} (2016) no.12,  121503.
  %%CITATION = doi:10.1103/PhysRevD.94.121503;%%

\bibitem{Yunes2011}
  N.~Yunes and L.~C.~Stein,
  %``Non-Spinning Black Holes in Alternative Theories of Gravity,''
  Phys.\ Rev.\ D {\bf 83} (2011) 104002.
  %%CITATION = doi:10.1103/PhysRevD.83.104002;%%
    
\bibitem{ABK} G. Antoniou, A. Bakopoulos and P. Kanti, 
  %``Evasion of No-Hair Theorems and Novel Black-Hole Solutions in Gauss-Bonnet Theories,''
  Phys.\ Rev.\ Lett.\  {\bf 120} (2018) no.13,  131102;
  %%CITATION = doi:10.1103/PhysRevLett.120.131102;%%
%``Black-Hole Solutions with Scalar Hair in Einstein-Scalar-Gauss-Bonnet Theories,''
  Phys.\ Rev.\ D {\bf 97} (2018) no.8,  084037.
  %%CITATION = doi:10.1103/PhysRevD.97.084037;%%

\bibitem{Doneva} D.~D.~Doneva and S.~S.~Yazadjiev,
  %``New Gauss-Bonnet Black Holes with Curvature-Induced Scalarization in Extended Scalar-Tensor Theories,''
  Phys.\ Rev.\ Lett.\  {\bf 120} (2018) no.13,  131103.
  %%CITATION = doi:10.1103/PhysRevLett.120.131103;%%

\bibitem{Silva} H.~O.~Silva, J.~Sakstein, L.~Gualtieri, T.~P.~Sotiriou and E.~Berti,
  %``Spontaneous scalarization of black holes and compact stars from a Gauss-Bonnet coupling,'
  Phys.\ Rev.\ Lett.\  {\bf 120} (2018) no.13,  131104.
  %%CITATION = doi:10.1103/PhysRevLett.120.131104;%%

\bibitem{Charmousis}
  C.~Charmousis, T.~Kolyvaris, E.~Papantonopoulos and M.~Tsoukalas,
  %``Black Holes in Bi-scalar Extensions of Horndeski Theories,''
  JHEP {\bf 1407} (2014) 085.
  %%CITATION = doi:10.1007/JHEP07(2014)085;%%

\bibitem{Correa}
  F.~Correa, M.~Hassaine and J.~Oliva,
  %``Black holes in New Massive Gravity dressed by a (non)minimally coupled scalar field,''
  Phys.\ Rev.\ D {\bf 89} (2014) no.12,  124005.
  %%CITATION = doi:10.1103/PhysRevD.89.124005;%%

\bibitem{Doneva-NS} D.~D.~Doneva and S.~S.~Yazadjiev,
  %``Neutron star solutions with curvature induced scalarization in the extended Gauss-Bonnet
%scalar-tensor theories,''
  JCAP {\bf 1804} (2018) no.04,  011.
  %%CITATION = doi:10.1088/1475-7516/2018/04/011;%%

\bibitem{Motohashi} H.~Motohashi and M.~Minamitsuji,
  %``General Relativity solutions in modified gravity,''
  Phys.\ Lett.\ B {\bf 781} (2018) 728;
  %%CITATION = doi:10.1016/j.physletb.2018.04.041;%%
  %``Stealth Schwarzschild solution in shift symmetry breaking theories,''
  Phys.\ Rev.\ D {\bf 98} (2018) no.8,  084027.
  %%CITATION = doi:10.1103/PhysRevD.98.084027;%%

\bibitem{Radu} C.~A.~R.~Herdeiro, E.~Radu, N.~Sanchis-Gual and J.~A.~Font,
  %``Spontaneous Scalarization of Charged Black Holes,''
  Phys.\ Rev.\ Lett.\  {\bf 121} (2018) no.10,  101102;
  %%CITATION = doi:10.1103/PhysRevLett.121.101102;%%
T.~Delsate, C.~Herdeiro and E.~Radu,
  %``Non-perturbative spinning black holes in dynamical Chern-Simons gravity,''
  Phys.\ Lett.\ B {\bf 787} (2018) 8;
  %%CITATION = doi:10.1016/j.physletb.2018.09.060;%%
  Y.~Brihaye, C.~Herdeiro and E.~Radu,
  %``The scalarised Schwarzschild-NUT spacetime,''
  Phys.\ Lett.\ B {\bf 788} (2019) 295.
%%CITATION = doi:10.1016/j.physletb.2018.11.022;%%

\bibitem{Doneva-Papa}
  D.~D.~Doneva, S.~Kiorpelidi, P.~G.~Nedkova, E.~Papantonopoulos and S.~S.~Yazadjiev,
  %``Charged Gauss-Bonnet black holes with curvature induced scalarization in the extended scalar-tensor theories,''
  Phys.\ Rev.\ D {\bf 98} (2018) no.10,  104056.
  %%CITATION = doi:10.1103/PhysRevD.98.104056;%%

\bibitem{Butler} M.~Butler, A.~M.~Ghezelbash, E.~Massaeli and M.~Motaharfar,
  %``Atiyah-Hitchin in Five Dimensional Einstein-Gauss-Bonnet Gravity,''
  arXiv:1808.03217 [hep-th].
  %%CITATION = ARXIV:1808.03217;%%

\bibitem{Danila}
  B.~Danila, T.~Harko, F.~S.~N.~Lobo and M.~K.~Mak,
  %``Spherically symmetric static vacuum solutions in hybrid metric-Palatini gravity,''
  arXiv:1811.02742 [gr-qc].
  %%CITATION = ARXIV:1811.02742;%%

\bibitem{Stetsko}
  M.~M.~Stetsko,
  %``Slowly rotating black hole solution in the scalar-tensor theory with nonminimal derivative coupling and its thermodynamics,''
  arXiv:1811.05030 [hep-th].
  %%CITATION = ARXIV:1811.05030;%%

\bibitem{Ayzen} K.~Yagi, L.~C.~Stein, N.~Yunes and T.~Tanaka,
  %``Post-Newtonian, Quasi-Circular Binary Inspirals in Quadratic Modified Gravity,''
 Phys.\ Rev.\ D {\bf 85} (2012) 064022
 Erratum: [Phys.\ Rev.\ D {\bf 93} (2016) no.2,  029902];
 %%CITATION = doi:10.1103/PhysRevD.93.029902, 10.1103/PhysRevD.85.064022;%%
  D.~Ayzenberg, K.~Yagi and N.~Yunes,
  %``Linear Stability Analysis of Dynamical Quadratic Gravity,''
  Phys.\ Rev.\ D {\bf 89} (2014) no.4,  044023.
  %%CITATION = doi:10.1103/PhysRevD.89.044023;%%

\bibitem{Dolan}
  S.~R.~Dolan, S.~Ponglertsakul and E.~Winstanley,
  %``Stability of black holes in Einstein-charged scalar field theory in a cavity,''
  Phys.\ Rev.\ D {\bf 92} (2015) no.12,  124047.
  %%CITATION = doi:10.1103/PhysRevD.92.124047;%%

\bibitem{Kunz} 
J.~L.~Blazquez-Salcedo, C.~F.~B.~Macedo, V.~Cardoso, V.~Ferrari, L.~Gualtieri, F.~S.~Khoo, J.~Kunz and P.~Pani,
  %``Perturbed black holes in Einstein-dilaton-Gauss-Bonnet gravity: Stability, ringdown,
  %and gravitational-wave emission,''
Phys.\ Rev.\ D {\bf 94} (2016) no.10,  104024.
   %%CITATION = doi:10.1103/PhysRevD.94.104024;%%

\bibitem{Bhatta} S.~Bhattacharya and S.~Chakraborty,
  %``Constraining some Horndeski gravity theories,''
  Phys.\ Rev.\ D {\bf 95} (2017) no.4,  044037;
   %%CITATION = doi:10.1103/PhysRevD.95.044037;%%
  I.~Banerjee, S.~Chakraborty and S.~SenGupta,
  %``Excavating black hole continuum spectrum: Possible signatures of scalar hairs and of higher dimensions,''
  Phys.\ Rev.\ D {\bf 96} (2017) no.8,  084035.
%%CITATION = doi:10.1103/PhysRevD.96.084035;%%

\bibitem{Tatter}
  O.~J.~Tattersall, P.~G.~Ferreira and M.~Lagos,
  %``Speed of gravitational waves and black hole hair,''
  Phys.\ Rev.\ D {\bf 97} (2018) no.8,  084005.
  %%CITATION = doi:10.1103/PhysRevD.97.084005;%%

\bibitem{Mukherjee}
S.~Mukherjee and S.~Chakraborty,
  %``Horndeski theories confront the Gravity Probe B experiment,''
  Phys.\ Rev.\ D {\bf 97} (2018) no.12,  124007.
  %%CITATION = doi:10.1103/PhysRevD.97.124007;%% 

\bibitem{Chakra}
S.~Chakrabarti,
  %``Collapsing spherical star in Scalar-Einstein-Gauss-Bonnet gravity with a quadratic coupling,''
  Eur.\ Phys.\ J.\ C {\bf 78} (2018) no.4,  296.
  %%CITATION = doi:10.1140/epjc/s10052-018-5798-9;%%

\bibitem{Berti} E.~Berti, K.~Yagi and N.~Yunes,
  %``Extreme Gravity Tests with Gravitational Waves from Compact Binary Coalescences: (I) Inspiral-Merger,''
  Gen.\ Rel.\ Grav.\  {\bf 50} (2018) no.4,  46.
  %%CITATION = doi:10.1007/s10714-018-2362-8;%%

\bibitem{Brihaye}
  Y.~Brihaye and B.~Hartmann,
  %``Critical phenomena of charged Einstein-Gauss-Bonnet black holes with charged scalar hair,''
  Class.\ Quant.\ Grav.\  {\bf 35} (2018) no.17,  175008.
  %%CITATION = doi:10.1088/1361-6382/aad389;%%

\bibitem{Prabhu} K.~Prabhu and L.~C.~Stein,
  %``Black hole scalar charge from a topological horizon integral in Einstein-dilaton-Gauss-Bonnet gravity,''
  Phys.\ Rev.\ D {\bf 98} (2018) no.2,  021503.
  %%CITATION = doi:10.1103/PhysRevD.98.021503;%%

\bibitem{Myung} Y.~S.~Myung and D.~C.~Zou,
  %``Gregory-Laflamme instability of black hole in Einstein-scalar-Gauss-Bonnet theories,''
  Phys.\ Rev.\ D {\bf 98} (2018) no.2,  024030;
  %%CITATION = doi:10.1103/PhysRevD.98.024030;%%
  %``Quasinormal modes of scalarized black holes in the Einstein-Maxwell-Scalar theory,''
  arXiv:1812.03604 [gr-qc].
  %%CITATION = ARXIV:1812.03604;%%

\bibitem{Don-Kunz}
  J.~L.~Blazquez-Salcedo, D.~D.~Doneva, J.~Kunz and S.~S.~Yazadjiev,
  %``Radial perturbations of the scalarized Einstein-Gauss-Bonnet black holes,''
  Phys.\ Rev.\ D {\bf 98} (2018) no.8,  084011;
  %%CITATION = doi:10.1103/PhysRevD.98.084011;%%
J.~L.~Blazquez-Salcedo, Z.~Altaha Motahar, D.~D.~Doneva, F.~S.~Khoo, J.~Kunz, S.~Mojica, K.~V.~Staykov and S.~S.~Yazadjiev,
  %``Quasinormal modes of compact objects in alternative theories of gravity,''
  arXiv:1810.09432 [gr-qc].
  %%CITATION = ARXIV:1810.09432;%%

\bibitem{Benkel2018}
  R.~Benkel, N.~Franchini, M.~Saravani and T.~P.~Sotiriou,
  %``Causal structure of black holes in shift-symmetric Horndeski theories,''
  Phys.\ Rev.\ D {\bf 98} (2018) no.6,  064006.
%%CITATION = doi:10.1103/PhysRevD.98.064006;%%

\bibitem{Iorio} L.~Iorio and M.~L.~Ruggiero,
  %``Constraining some $r^{-n}$ extra-potentials in modified gravity models with LAGEOS-type laser-ranged geodetic satellites,''
  JCAP {\bf 1810} (2018) no.10,  021.
%%CITATION = doi:10.1088/1475-7516/2018/10/021;%%

\bibitem{Ovalle}
  J.~Ovalle, R.~Casadio, R.~da Rocha, A.~Sotomayor and Z.~Stuchlik,
  %``Einstein-Klein-Gordon system by gravitational decoupling,''
  EPL {\bf 124} (2018) no.2,  20004;
  %%CITATION = doi:10.1209/0295-5075/124/20004;%%}
  J.~Ovalle,
  %``Decoupling gravitational sources in general relativity: The extended case,''
  Phys.\ Lett.\ B {\bf 788} (2019) 213.
  %%CITATION = doi:10.1016/j.physletb.2018.11.029;%%

\bibitem{Barack}
  L.~Barack {\it et al.},
  %``Black holes, gravitational waves and fundamental physics: a roadmap,''
  arXiv:1806.05195 [gr-qc].
  %%CITATION = ARXIV:1806.05195;%%

\bibitem{Gao} Y.~X.~Gao, Y.~Huang and D.~J.~Liu,
  %``Scalar perturbations on the background of Kerr black holes in the quadratic dynamical Chern-Simons gravity,''
  arXiv:1808.01433 [gr-qc].
  %%CITATION = ARXIV:1808.01433;%%

\bibitem{Lee} B.~H.~Lee, W.~Lee and D.~Ro,
  %``Expanded evasion of the black hole no-hair theorem in dilatonic Einstein-Gauss-Bonnet theory,''
  arXiv:1809.05653 [gr-qc].
  %%CITATION = ARXIV:1809.05653;%%

\bibitem{Witek2018}
  H.~Witek, L.~Gualtieri, P.~Pani and T.~P.~Sotiriou,
  %``Black holes and binary mergers in scalar Gauss-Bonnet gravity: scalar field dynamics,''
  arXiv:1810.05177 [gr-qc].
  %%CITATION = ARXIV:1810.05177;%%

\bibitem{Moto} H.~Motohashi and S.~Mukohyama,
  %``Shape dependence of spontaneous scalarization,''
  arXiv:1810.12691 [gr-qc].
  %%CITATION = ARXIV:1810.12691;%%

\bibitem{Narita}
  T.~Torii, K.~Maeda and M.~Narita,
  %``No scalar hair conjecture in asymptotic de Sitter space-time,''
  Phys.\ Rev.\ D {\bf 59} (1999) 064027.
  %%CITATION = doi:10.1103/PhysRevD.59.064027;%%

\bibitem{Winstanley-Nohair} E.~Winstanley,
  %``On the existence of conformally coupled scalar field hair for black holes in (anti-)de Sitter space,''
  Found.\ Phys.\  {\bf 33} (2003) 111;
  %%CITATION = doi:10.1023/A:1022871809835;%%
  %``Dressing a black hole with non-minimally coupled scalar field hair,''
  Class.\ Quant.\ Grav.\  {\bf 22} (2005) 2233.
  %%CITATION = doi:10.1088/0264-9381/22/11/020;%%

\bibitem{Bhatta2007}
  S.~Bhattacharya and A.~Lahiri,
  %``Black-hole no-hair theorems for a positive cosmological constant,''
  Phys.\ Rev.\ Lett.\  {\bf 99} (2007) 201101.
  %%CITATION = doi:10.1103/PhysRevLett.99.201101;%%

\bibitem{Bardoux}
  Y.~Bardoux, M.~M.~Caldarelli and C.~Charmousis,
  %``Shaping black holes with free fields,''
  JHEP {\bf 1205} (2012) 054
  doi:10.1007/JHEP05(2012)054
  [arXiv:1202.4458 [hep-th]].
  %%CITATION = doi:10.1007/JHEP05(2012)054;%%

\bibitem{Kazanas}
  J.~Sultana and D.~Kazanas,
  %``A no-hair theorem for spherically symmetric black holes in $R^2$ gravity,''
  Gen.\ Rel.\ Grav.\  {\bf 50} (2018) no.11,  137.
 %%CITATION = doi:10.1007/s10714-018-2463-4;%%

\bibitem{Martinez-deSitter}
  C.~Martinez, R.~Troncoso and J.~Zanelli,
  %``De Sitter black hole with a conformally coupled scalar field in four-dimensions,''
  Phys.\ Rev.\ D {\bf 67} (2003) 024008.
  %%CITATION = doi:10.1103/PhysRevD.67.024008;%%

\bibitem{Harper}
  T.~J.~T.~Harper, P.~A.~Thomas, E.~Winstanley and P.~M.~Young,
  %``Instability of a four-dimensional de Sitter black hole with a conformally coupled scalar field,''
  Phys.\ Rev.\ D {\bf 70} (2004) 064023.
  %%CITATION = doi:10.1103/PhysRevD.70.064023;%%

\bibitem{Martinez} M.~Henneaux, C.~Martinez, R.~Troncoso and J.~Zanelli,
  %``Asymptotically anti-de Sitter spacetimes and scalar fields with a logarithmic branch,''
  Phys.\ Rev.\ D {\bf 70} (2004) 044034;
  %%CITATION = doi:10.1103/PhysRevD.70.044034;%%
  C.~Martinez, R.~Troncoso and J.~Zanelli,
  %``Exact black hole solution with a minimally coupled scalar field,''
  Phys.\ Rev.\ D {\bf 70} (2004) 084035;
  %%CITATION = doi:10.1103/PhysRevD.70.084035;%%
  C.~Erices and C.~Martinez,
  %``Rotating hairy black holes in arbitrary dimensions,''
  Phys.\ Rev.\ D {\bf 97} (2018) no.2,  024034.
  %%CITATION = doi:10.1103/PhysRevD.97.024034;%%

\bibitem{Radu-Win}
  E.~Radu and E.~Winstanley,
  %``Conformally coupled scalar solitons and black holes with negative cosmological constant,''
  Phys.\ Rev.\ D {\bf 72} (2005) 024017.
  %%CITATION = doi:10.1103/PhysRevD.72.024017;%%

\bibitem{Anabalon}
  A.~Anabalon and H.~Maeda,
  %``New Charged Black Holes with Conformal Scalar Hair,''
  Phys.\ Rev.\ D {\bf 81} (2010) 041501.
  %%CITATION = doi:10.1103/PhysRevD.81.041501;%%

\bibitem{Hosler}
  D.~Hosler and E.~Winstanley,
  %``Higher-dimensional solitons and black holes with a non-minimally coupled scalar field,''
  Phys.\ Rev.\ D {\bf 80} (2009) 104010.
  %%CITATION = doi:10.1103/PhysRevD.80.104010;%%

\bibitem{Kolyvaris} C.~Charmousis, T.~Kolyvaris and E.~Papantonopoulos,
  %``Charged C-metric with conformally coupled scalar field,''
  Class.\ Quant.\ Grav.\  {\bf 26} (2009) 175012;
  %%CITATION = doi:10.1088/0264-9381/26/17/175012;%%
  T.~Kolyvaris, G.~Koutsoumbas, E.~Papantonopoulos and G.~Siopsis,
  %``A New Class of Exact Hairy Black Hole Solutions,''
  Gen.\ Rel.\ Grav.\  {\bf 43} (2011) 163.
  %%CITATION = doi:10.1007/s10714-010-1079-0;%%

\bibitem{Ohta}
  K.~i.~Maeda, N.~Ohta and Y.~Sasagawa,
  %``AdS Black Hole Solution in Dilatonic Einstein-Gauss-Bonnet Gravity,''
  Phys.\ Rev.\ D {\bf 83} (2011) 044051;
  %%CITATION = doi:10.1103/PhysRevD.83.044051;%%
Z.~K.~Guo, N.~Ohta and T.~Torii,
  %``Black Holes in the Dilatonic Einstein-Gauss-Bonnet Theory in Various Dimensions II. Asymptotically AdS Topological Black Holes,''
  Prog.\ Theor.\ Phys.\  {\bf 121} (2009) 253;
  %%CITATION = doi:10.1143/PTP.121.253;%%
N.~Ohta and T.~Torii,
  %``Black Holes in the Dilatonic Einstein-Gauss-Bonnet Theory in Various Dimensions. III. Asymptotically AdS Black Holes with k = +-1,''
  Prog.\ Theor.\ Phys.\  {\bf 121} (2009) 959; 
  %%CITATION = doi:10.1143/PTP.121.959;%%
 N.~Ohta and T.~Torii,
  %``Black Holes in the Dilatonic Einstein-Gauss-Bonnet Theory in Various Dimensions IV: Topological Black Holes with and without Cosmological Term,''
  Prog.\ Theor.\ Phys.\  {\bf 122} (2009) 1477.
%%CITATION = doi:10.1143/PTP.122.1477;%%

\bibitem{Saenz}
  S.~G.~Saenz and C.~Martinez,
  %``Anti-de Sitter massless scalar field spacetimes in arbitrary dimensions,''
  Phys.\ Rev.\ D {\bf 85} (2012) 104047.
  %%CITATION = doi:10.1103/PhysRevD.85.104047;%%

\bibitem{Caldarelli}
  M.~M.~Caldarelli, C.~Charmousis and M.~Hassaine,
  %``AdS black holes with arbitrary scalar coupling,''
  JHEP {\bf 1310} (2013) 015.
  %%CITATION = doi:10.1007/JHEP10(2013)015;%%

\bibitem{Gonzalez}
  P.~A.~Gonzalez, E.~Papantonopoulos, J.~Saavedra and Y.~Vasquez,
  %``Four-Dimensional Asymptotically AdS Black Holes with Scalar Hair,''
  JHEP {\bf 1312} (2013) 021.
  %%CITATION = doi:10.1007/JHEP12(2013)021;%%

\bibitem{Gaete}
M.~Bravo Gaete and M.~Hassaine,
  %``Topological black holes for Einstein-Gauss-Bonnet gravity with a nonminimal scalar field,''
  Phys.\ Rev.\ D {\bf 88} (2013) 104011;
  %%CITATION = doi:10.1103/PhysRevD.88.104011;%%
  M.~Bravo Gaete and M.~Hassaine,
  %``Planar AdS black holes in Lovelock gravity with a nonminimal scalar field,''
  JHEP {\bf 1311} (2013) 177.
  %%CITATION = doi:10.1007/JHEP11(2013)177;%%

\bibitem{Giribet}
  G.~Giribet, M.~Leoni, J.~Oliva and S.~Ray,
  %``Hairy black holes sourced by a conformally coupled scalar field in D dimensions,''
  Phys.\ Rev.\ D {\bf 89} (2014) no.8,  085040.
  %%CITATION = doi:10.1103/PhysRevD.89.085040;%%

\bibitem{BenAchour}
  J.~Ben Achour and H.~Liu,
  %``Stealth Schwarzschild-(A)dS black hole in DHOST theories after GW170817: Linear
  %time-dependent scalar dressing,''
  arXiv:1811.05369 [gr-qc].
  %%CITATION = ARXIV:1811.05369;%%

\bibitem{Hartmann} Y.~Brihaye, B.~Hartmann and J.~Urrestilla,
  %``Solitons and black hole in shift symmetric scalar-tensor gravity with cosmological constant,''
  JHEP {\bf 1806} (2018) 074;
  %%CITATION = doi:10.1007/JHEP06(2018)074;%%
  Y.~Brihaye and B.~Hartmann,
  %``Charged scalar-tensor solitons and black holes with (approximate) Anti-de Sitter asymptotics,''
  arXiv:1810.05108 [gr-qc].
  %%CITATION = ARXIV:1810.05108;%%
  
\bibitem{GH} G.~W.~Gibbons and S.~W.~Hawking,
  %``Action Integrals and Partition Functions in Quantum Gravity,''
  Phys.\ Rev.\ D {\bf 15} (1977) 2752.
  %%CITATION = doi:10.1103/PhysRevD.15.2752;%%
  
\bibitem{York} J.~W.~York, Jr.,
  %``Black Hole in Thermal Equilibrium With a Scalar Field: The Back Reaction,''
  Phys.\ Rev.\ D {\bf 31} (1985) 775.
  %%CITATION = doi:10.1103/PhysRevD.31.775;%%
  
\bibitem{GK} G.~W.~Gibbons and R.~E.~Kallosh,
  %``Topology, entropy and Witten index of dilaton black holes,''
  Phys.\ Rev.\ D {\bf 51} (1995) 2839.
  %%CITATION = doi:10.1103/PhysRevD.51.2839;%% 

\bibitem{BAK2} A. Bakopoulos. G. Antoniou and P. Kanti, in progress.

\bibitem{HP} S.~W.~Hawking and D.~N.~Page,
  %``Thermodynamics of Black Holes in anti-De Sitter Space,''
  Commun.\ Math.\ Phys.\  {\bf 87} (1983) 577.
  %%CITATION = doi:10.1007/BF01208266;%%\\

\bibitem{Wald}
  R.~M.~Wald,
  %``Black hole entropy is the Noether charge,''
  Phys.\ Rev.\ D {\bf 48} (1993) no.8,  R3427.
  %%CITATION = doi:10.1103/PhysRevD.48.R3427;%%

\bibitem{Dutta}
  S.~Dutta and R.~Gopakumar,
  %``On Euclidean and Noetherian entropies in AdS space,''
  Phys.\ Rev.\ D {\bf 74} (2006) 044007.
  %%CITATION = doi:10.1103/PhysRevD.74.044007;%%

\bibitem{Iyer}
  V.~Iyer and R.~M.~Wald,
  %``Some properties of Noether charge and a proposal for dynamical black hole entropy,''
  Phys.\ Rev.\ D {\bf 50} (1994) 846.
  %%CITATION = doi:10.1103/PhysRevD.50.846;%%

\end{thebibliography}
\end{document}